# The Quantum Technology Job Market: Data Driven Analysis of 3641 Job Posts

Simon Goorney (1,2,3), Eleni Karydi (1,2,3), Borja Muñoz (1,3), Otto Santesson (1), Zeki Can Seskir (4), Ana Alina Tudoran (1), Jacob Sherson (1,2,3)*.

*1: Department of Management, School of Business and Social Science,*

*Aarhus University, Aarhus, Denmark*

*2: Niels Bohr Institute, Copenhagen University, Copenhagen, Denmark*

*3: European Quantum Readiness Center, Aarhus, Denmark*

*4: Karlsruhe Institute of Technology - Institute for Technology Assessment and*

*Systems Analysis, Karlsruhe, Germany*

**sherson@mgmt.au.dk*

## Abstract

The rapid advancement of Quantum Technology (QT) has created a growing demand for a specialized workforce, spanning across academia and industry. This study presents a quantitative analysis of the QT job market by systematically extracting and classifying 3641 job postings worldwide. The classification pipeline leverages large language models (LLMs) whilst incorporating a "human-in-the-loop" validation process to ensure reliability. The research identifies key trends in regional job distribution, degree and skill requirements, and the evolving demand for QT-related roles. Findings reveal a strong presence of the QT job market in the United States and Europe, with increasing corporate demand for engineers, software developers, and PhD-level researchers. Despite growing industry applications, the sector remains in its early stages, dominated by large technology firms. However, there is some sign that there may be change to come, indicated by a strong presence of Bachelor and Master's graduates in the workforce.

## Keywords

## Abbreviations

CFQT: Competence Framework in Quantum Technologies

DigiQ: Digitally Enhanced Quantum Technology Master

EU: European Union

HQ: Headquarters

ICR: Intercoder Reliability

LLMs: Language Large Models

QT: Quantum Technology

QTEdu: Quantum Technology Education

QTs: Quantum Technologies

R&D: Research and Development

RTO: Research and Technology Organisation

STEM: Science, Technology, Engineering and Mathematics

UK: United Kingdom

USA: United States of America

VC: Venture Capital

# 1. Introduction

The potential applications of QT, in particular quantum computing, have been demonstrated in areas such as cybersecurity, materials and pharmaceuticals, banking and finance and advanced manufacturing [1]. Despite this progress, the commercial applications and maturity of QT, require "significant investment, expertise and cultural change" [2]. Due to its potential impact to revolutionise society and economy, a quantum-ready ecosystem is needed to accelerate advancements in the emerging quantum industry, with strategic decisions from governments worldwide [2].

Industrial and national assessments have highlighted the need for standards and benchmarking of new developments [3], greater international collaboration allowing free international trade, increasing awareness of the value of quantum technologies over traditional solutions, and a stronger pipeline for workforce development to provide talent to the industry worldwide [4]. Interdisciplinary jobs in QT are increasing in number [5-7] because supplying the industry with a highly skilled workforce and talent pipeline is an important concern. This includes a workforce composed not only of PhD graduates, but of bachelor and master graduates from many different backgrounds, as QT becomes increasingly diversified [5].

The required characteristics of the workforce have been investigated in previous studies [5-13]. One of the main challenges addressed is the "workforce gap" as more QTs are progressively moving from theory and lab experimentation towards commercial applications [8] and are expected to gain relevance in the future [9]. In overcoming that gap, it is essential to understand, which are the competencies and skills needed for entering the quantum industry [8-9, 11, 13-16] and to serve as a foundational guide for future educational programs [17-18].

Shedding light on the most important skills for the quantum industry, from an industrial perspective, Fox et al. [11], conducted interviews with 21 companies, and reported that 90% of them considered coding skills and statistical methods for data analysis to be highly valuable, followed by laboratory experience (81%), electronics (76%), troubleshooting and problem solving (71%), knowledge of material science and properties (67%) and quantum algorithms and computer science (62%) [11].

Apart from mastering technical skills, the need for “enterprise skills” or soft business skills is growing across suppliers and end users of QT [15]. This trend aligns with the progressive commercialization of QTs, requiring diverse profiles to transmit the value and potential use cases of QTs. However, the skills needed may not be the same for every role, as stated by Hughes et al. [5]. In their analysis, they identified 3 distinct skills clusters (hardware, software and business), valuable to industry. It is notable that these clusters are not necessarily overlapping, and universities could structure their programs to align with one or more of them to better prepare students for specialised roles in the quantum industry. The authors also suggested that personnel with a non-quantum background should be encouraged to enter the quantum industry in order to diversify the pool of skills available to employers [5] and to manage the shortage of trained talent. For example, upskilling programs have been effective for employing photonic technicians [8].Similarly, a recent study from Greinert et al. [6] with 34 interviews coming from a wide range of companies, identified the need for training programs to be formed based on different projected job roles, from engineers or technicians requiring technical skills to marketing and sales roles with a greater holistic understanding of QT and the market [6].

Hands-on experience is undoubtedly a great asset for entering the quantum industry, as reported by companies [5-13], a trend that has been observed both in interview studies and in the structure of educational programs [8]. Hence, the quantum educational landscape is evolving rapidly with an increasing diversity of opportunities emerging, from certificates [19-22], minors [23-26], industry training [27], educational resources [12], to different training possibilities across sectors [6].

As for the most desired roles for the quantum industry, there is a distinction depending on the level of specialisation and skills, ranging from specialists focusing on research and development of new devices or theory, to non-quantum engineers (with increasing number of job positions) employed in areas such as electronics, software, or sales [17]. Higher levels of specialisation may likely require a PhD, while master students or bachelors will contribute to the workforce pipeline in roles demanding less specialisation [17].

Fox et al., identified engineering roles as the most desired ones (95% of surveyed companies), although within these roles there may also be differences between companies in the specific skills required [11]. In the same way, Hughes et al. reported growing demand for engineering (95%) but a high proportion of experimentation and research, with roles such as “experimental physicist” (86%) and “theorists” (56%) still needed in the following years [5].

Finally, in a later study, Greinert et al suggested that we could expect a shifting from hardware to software roles as the industry matures [6].

An important development toward improving communication between industry and education programs is the Competence Framework in Quantum Technologies (CFQT) [16]. The CFQT is a continuous effort in mapping the landscape of skills in QT, aiming to be a comprehensive guide covering topic areas (content domains of the framework) which may appear in education programs, and competences which may be required for certain jobs. It has been developed over several years within the EU, incorporating multiple rounds of stakeholder interviews and community feedback. While the first versions comprised primarily a topic map, later versions have included proficiency levels [28],which can be mapped to educational program levels, and qualification profiles, which can frame the job roles in terms of their educational needs. While as yet the CFQT has not reached complete adoption, it is beginning to appear in educational programs throughout the EU [27,29]. The extent to which it represents the day-to-day of jobs, however, is less clear, and this can only be established by a large-scale study of the skills involved in the industry. This research is beyond the scope of the present article, however this will likely be a fruitful direction for future work.

In summary, the need for a skilled quantum workforce is now well known. However, the best route to this workforce is not. It is of paramount importance to develop more education programs worldwide. But at what level? And what job roles should the graduates be going into? Answers to these questions are needed in order to smoothly address the growing issue of talent worldwide.

There is a wide variety of different approaches to industry development worldwide. Given this variance, it is crucial and timely to ask what exactly is the current state of the quantum industry, in which the graduates of the rapidly expanding education programs will likely find work in the future. Engineering and technical roles, in particular, have been highlighted by prior studies [5,6,11], as being important for the development of QT in the coming years. But to what extent is that reflected in the job market? In this article, we are investigating in a quantitative manner, the status and needs of the quantum industry, using data from thousands of job posts in QT. This work represents a step toward improving our understanding of how far in development QT is, how close it is to commercialising , and what kind of workforce it would need to boost it forward. Furthermore, this work introduces an innovative classification scheme, making use of large language models (LLMs) and combining them with human expert labelling, to generate a sizable dataset of over 3600 job posts, shedding light on the quantum labour market worldwide.

## 2. Research questions

In this research, we attempt to elucidate the reality of the industry needs and identify the actions required to address them using quantitative methodology. Over the course of 2023 and 2024, we have gathered job posts relating to QT and in this article we use this large-scale job market data to understand the worldwide landscape of QT, through specific research questions:

RQ1: What are the characteristics of the QT Job marketplace worldwide?

RQ2: Are there regional differences between the QT job marketplace worldwide, particularly between the EU and the USA?

RQ3: What are the educational requirements of the QT workforce?

The backbone of this research has been the development of a robust algorithm to extract jobs from online platforms and categorise them in a multi-step classification pipeline. The next section details this process.

# 3. Methodology

Historically, classification of qualitative data has been limited to human coders [30,31] often in the form of research assistants or through services such as Amazon's Mechanical Turk [32]. The process requires significant time and resources and it is, moreover, liable to inconsistencies stemming from individual biases and variability in interpretation of the codebook. Another drawback to human classification is its inherent incompatibility with scalability, making the classification of a large body of data almost impossible. Recent advancements in large language models (LLMs), particularly pre-trained models such as ChatGPT-4o-mini [33], represent a paradigm shift in the classification and analysis of qualitative data. These models mitigate many of the above mentioned limitations associated with human coders by leveraging their computational efficiency, consistency, and scalability [34-36].

In the context of the investigation of the quantum technology job market, these capabilities are especially advantageous. LLMs offer an innovative methodology by transforming qualitative data—such as job postings and descriptions—into structured, analyzable formats. This transformation enables the application of data-centric approaches to derive insights that were previously elusive; by utilizing the vast training data encoded in a pre-trained LLM, combined with a human-in-the-loop methodology [37-38] these models can identify key skills, emerging roles, and other key pointers within the quantum job market. Another key advantage of using LLMs for classification is their ability to rapidly label data from large, dynamic datasets. This makes them particularly well-suited for continuously retrieving and analyzing job postings from online job boards. By handling the constant influx of new data, LLMs enable real-time insights into the evolving trends of the quantum technology job market, a capability that traditional methods struggle to match. For details around the human validation process, see Appendix 1.

## 3.1 Classification steps

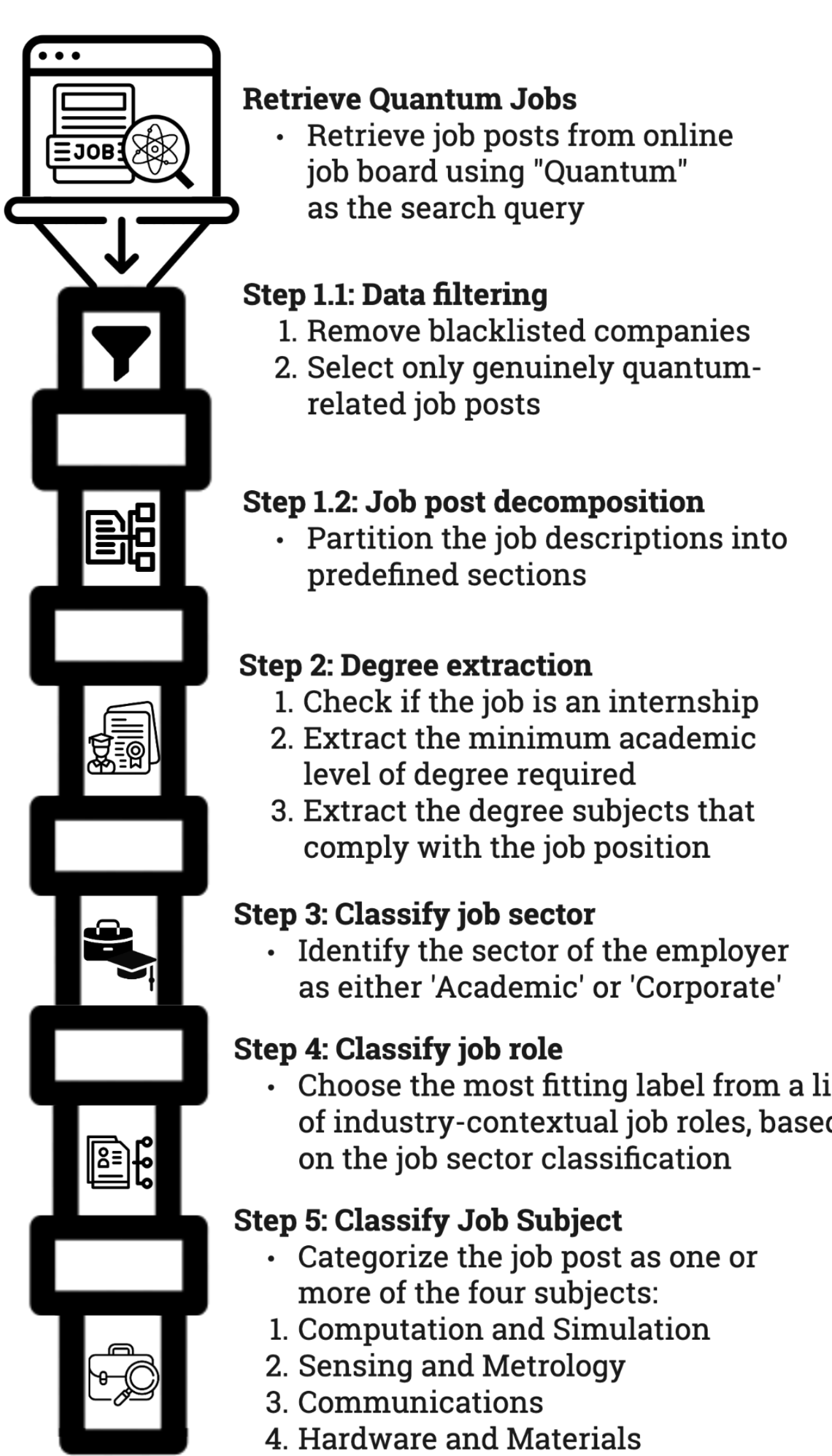

Figure 1. Quantum job classification process. The figure illustrates the methodology used for data extraction and the steps conducted for classification across multiple categories. The process begins by extracting quantum jobs from an online database (step 1) and identifying the degree requirements for each one (step 2). After that, three classifications follow (steps 3-5) with the categories being: job sector, job role and job subjects. Each step of the process was validated by human experts, see Appendix 1.

### *Step 1: Retrieve Quantum jobs*

We used gpt-4o-mini-2024-07-18 [33] to classify quantum job postings from constructed prompts. This specific gpt model was used for all the steps described below, except from step 1.2, where gpt-4o-2024-08-06 was utilized instead and the validations of steps 1-3 that were initially performed with gpt-3.5-turbo-0125. The project relies on the data of job postings on an international labor market platform. The data was collected weekly over the time period 03/2023 to 06/2024, using the search query “quantum” and setting the location to “Worldwide”. The data was exported to an Excel file with the extraction date. After the data was saved in the project repository, an auxiliary script written in Python [39] retrieved additional data to the job postings, such as job description and employer information. The data is now ready to enter the first classification step: filtering and job post decomposition (see Appendix 1). These compose cascading LLM-based classification steps, which are produced and validated by human experts. The human-in-the-loop aspect of the pipeline is crucial in order to ensure reliable, representative data, something that will be touched upon later.

The data filtering step (1.1) starts out by removing blacklisted companies; companies that are known from experience of experts to not be associated with QT. It then proceeds to select only genuinely quantum-related job posts, using the LLM to classify each one as either quantum-related or not, based on a human-generated prompt. Duplicates are handled by removing multiple instances of rows that have the same Job ID (the ID column), since each job posting is paired with a unique job ID, provided by the job posting site. Detailed information around the prompts used can be found in Appendix 4.
The next substep is 1.2, job post decomposition, which partitions the job descriptions into predefined sections that may be used as a whole or selectively for the subsequent steps of the pipeline. This is useful, since it offers the possibility of inputting to subsequent steps only sections relevant to the specific LLM-classification, which saves computational resources. The decomposition results in the following nine sections: Organisation Information, Job Summary or Overview, Key Responsibilities and Duties, Required Skills, Preferred Skills, Required Qualifications, Preferred Qualifications, Required Previous Experience, Preferred Previous Experience. After step 1, the now only quantum-related job posts in the dataset are also stored with their job description decomposition. It should be noted that job descriptions are pulled from a single online platform only. This was decided by investigating the four most prevalent online platforms and identifying which jobs were repeated across different platforms. One single platform was demonstrably more comprehensive than the others, and therefore it was decided to use this single source, which may represent around 75% of all QT jobs available to view. See Appendix 2 for additional information.

***Step 2: Degree extraction***
The purpose of step 2 is to extract information about the degree required for the job position. This is achieved in three different substeps; the first one, 2.1, checks whether the position is an internship through pattern matching – if so, then 'Internship' is the classification label. The residual unlabelled job posts then continue into substep 2.2, which by pattern-matching assigns the minimum degree level, choosing among  the preselected labels: "Bachelor", "Master", "PhD" and "None specified", totalling thus five possible classifications, including "Internship" from the previous step.
The last substep, 2.3, extracts the degree subjects that are mentioned in the job position. For example, a job position for a quantum hardware engineer could have required that the degree must be in a relevant field, such as electronic engineering and quantum physics. The degrees mentioned are identified to a set of subjects by pattern matching. One job post can be associated with multiple degree domains.

***Step 3: Classify job sector***
Step 3 is concerned with the  job sector of the job samples, and the possible categories they could fit into, are  'Academic' or 'Corporate'. The decision is based on  information regarding the employer; that is, if the employer is a public, research institution, such as a university, or a research and technology organisation (RTO), the job will be classified as "Academic", while f the employer is a  private entity, the job will be labeled as "Corporate". The classification process starts by identifying the obvious instances using pattern matching, for example searching for keywords such as “national lab” or “university”. Afterwards, classification by the LLM is performed on the remaining job posts, until all of them have been characterized with one of the two labels.

***Step 4: Classify job role***
In step 4, the job role of the job post is identified through three substeps. Firstly, in substep 4.1, pattern-matching techniques are applied to the job title to identify clear instances where the job role is a PhD position. In substep 4.2, the LLM is employed to classify job roles based on job roles which are clustered from the whole dataset, and their associated descriptions: one for 'Academic' roles and another for 'Corporate' roles. For example, if a job post is labeled as 'Academic' in the previous step, it is mapped to the most suitable role within this category. The classification process is further refined in substep 4.3, where certain roles are re-evaluated to ensure high confidence in classifications. This is accomplished using a refined prompt to specifically address roles that are prone to misclassification, such as distinguishing between 'Postdoc' and 'Professor,' or 'Product Manager' and 'Project Manager.' By leveraging sector-specific labels, the potential range of choices available to the language model is reduced, thereby enhancing classification accuracy.

***Step 5: Classify Quantum domain***
Finally, the job posts enter step 5, the last classification step of the pipeline. In this step the LLM checks the responsibilities of each position described in the job posts and determines whether they fit into one or more of the four Quantum domains; Computation and Simulation, Sensing and Metrology, Communications, and Hardware and Materials. It is a multi-label classification, meaning that each job post can belong to several Quantum domains. The classification is carried out by letting the LLM extract areas of the QT field the responsibilities of the job are associated with. Pattern-matching is then performed on the extractions, resulting in the assignment of pillar-relating labels for each post. The patterns are generated through an iterative process of identifying which keywords are associated with which domains, running the sub-step, and refining the patterns until the result is satisfactory.

As described above, each step is performed in a successive manner, making the pipeline cumulative and semi-cascading, whereby all subsequent steps depend on the initial two filtering and decomposition steps, but not necessarily on all preceding steps. Only selected steps incorporate outputs from earlier stages as additional inputs, such as step 4.
The result of steps 1-5 is a detailed corpus of information on the quantum job marketplace. In this research, we used 3641 job posts and all steps were thoroughly validated by experts in QT and data scientists to ensure accuracy of classification. This involved manual classification of a subset of job posts (10%) and comparing human annotated results to those from the classification, as a form of inter coder reliability (ICR). A detailed overview of the validation process can be found in Appendix 1.

# 4. Results

Over the time period 03/2023 to 06/2024, a total of 3641 jobs were extracted from an online job board. In this section, we present the descriptive information available from classification steps 1-5, namely the regional distribution of the job posts (subsection 4.1), the degree level requirements (subsection 4.2), the degree subject requirements (subsection 4.3), the job sector (subsection 4.4), the job subjects classified by QT domains (subsection 4.5), and the job roles (subsection 4.6). Note that while 3641 comprises the total dataset of individual jobs, the total count shown on each plot differs from this number for several reasons, such as co-counted jobs (e.g in subsection 4.3) and the exclusion of jobs posted by recruitment companies (Fig. 3 and 4).

## 4.1 Regional distribution of QT jobs

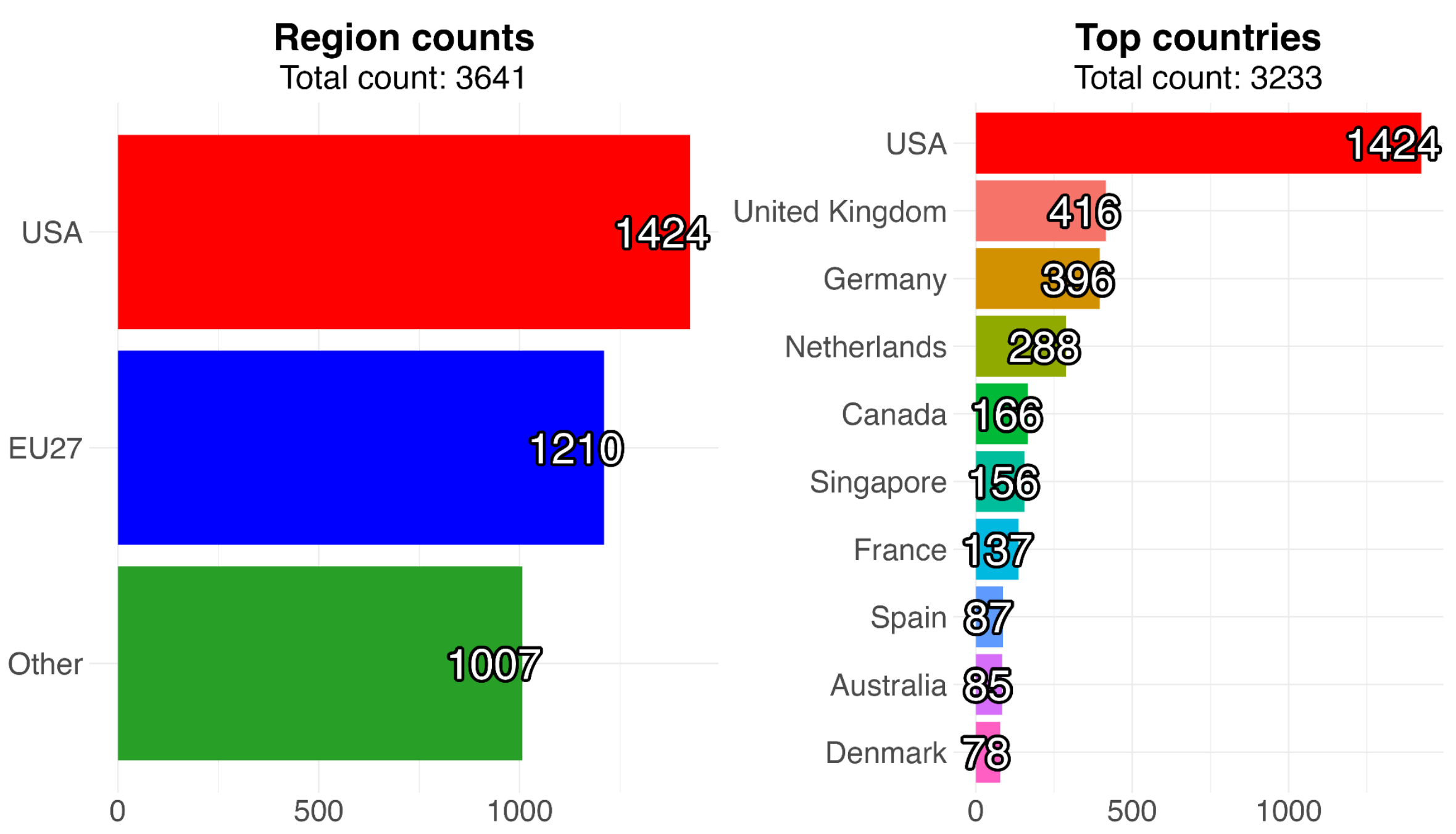

Figure 2. Number of countries posting quantum jobs worldwide. The figures show the distribution of quantum jobs globally with the left presenting the number of jobs found in the USA, in EU27 collectively and in all other countries, in order to directly compare between EU and USA. On the right figure, the distribution of QT jobs per country is presented, with the United States standing out, highlighting their dominant role in the quantum global market.

The USA has the highest count of quantum job postings (1424), while the EU27 has fewer but still a comparable number (1210). The dominance of the USA suggests that policy initiatives such as the National Quantum Initiative [40] have been effective at fostering a conducive environment for job creation in this sector. The EU, meanwhile, benefits from both a region-wide program, the Quantum Flagship [41], and country-specific funding.The UK, where the national quantum technology program [42] has been in place since 2013, has been an early adopter in QT and this may have contributed to the overall large number of jobs there compared to the size of the nation (the UK has a smaller population [416 jobs, 69 million population] than Germany [396 jobs, 84 million population]). The Netherlands also have a strong quantum ecosystem in comparison to the country's size (298 jobs, 18 million population), perhaps owing to the efforts of the Quantum Delta NL [43] which has been a significant boon to the Dutch quantum ecosystem.

Regarding how different sized companies are distributed worldwide, it is evident in Fig. 3 that overall, the majority of job positions are offered by large companies, with over 10,000 employees. We can glean further information from comparing this type of distribution between the EU27, USA, and other countries (Fig. 4). Note that in these plots, job posts advertised by recruitment companies are filtered out, as they do not accurately represent the company demographics. This reduces the total count from 3641 to 3037.

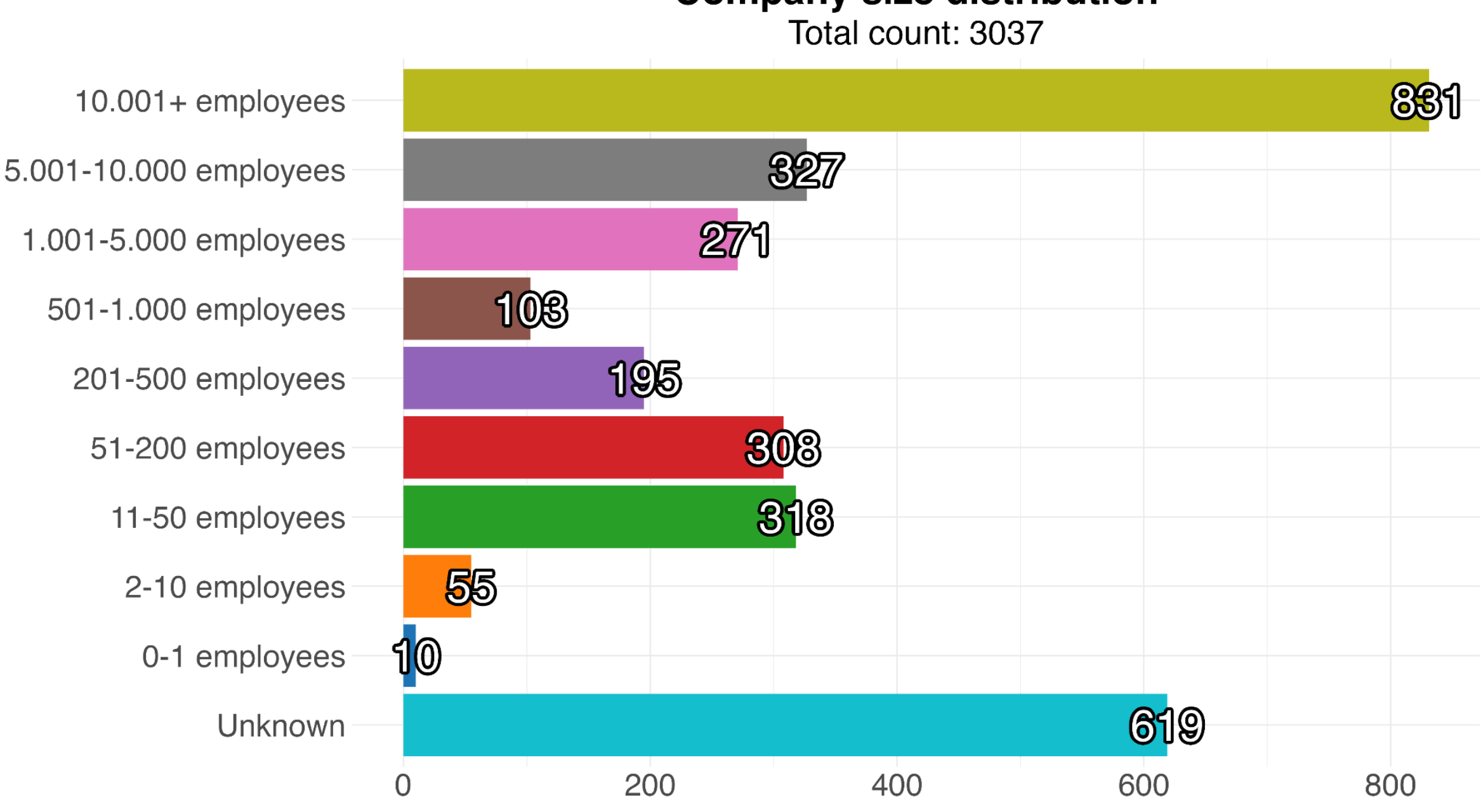

Figure 3. Company size distribution. The figure presents the number of job posts found in each company-size category, ranging from the smallest companies with 0-1 employees to the largest, employing more than 10000 people. The distribution shows large companies posting the most jobs over the duration of the dataset accumulation.

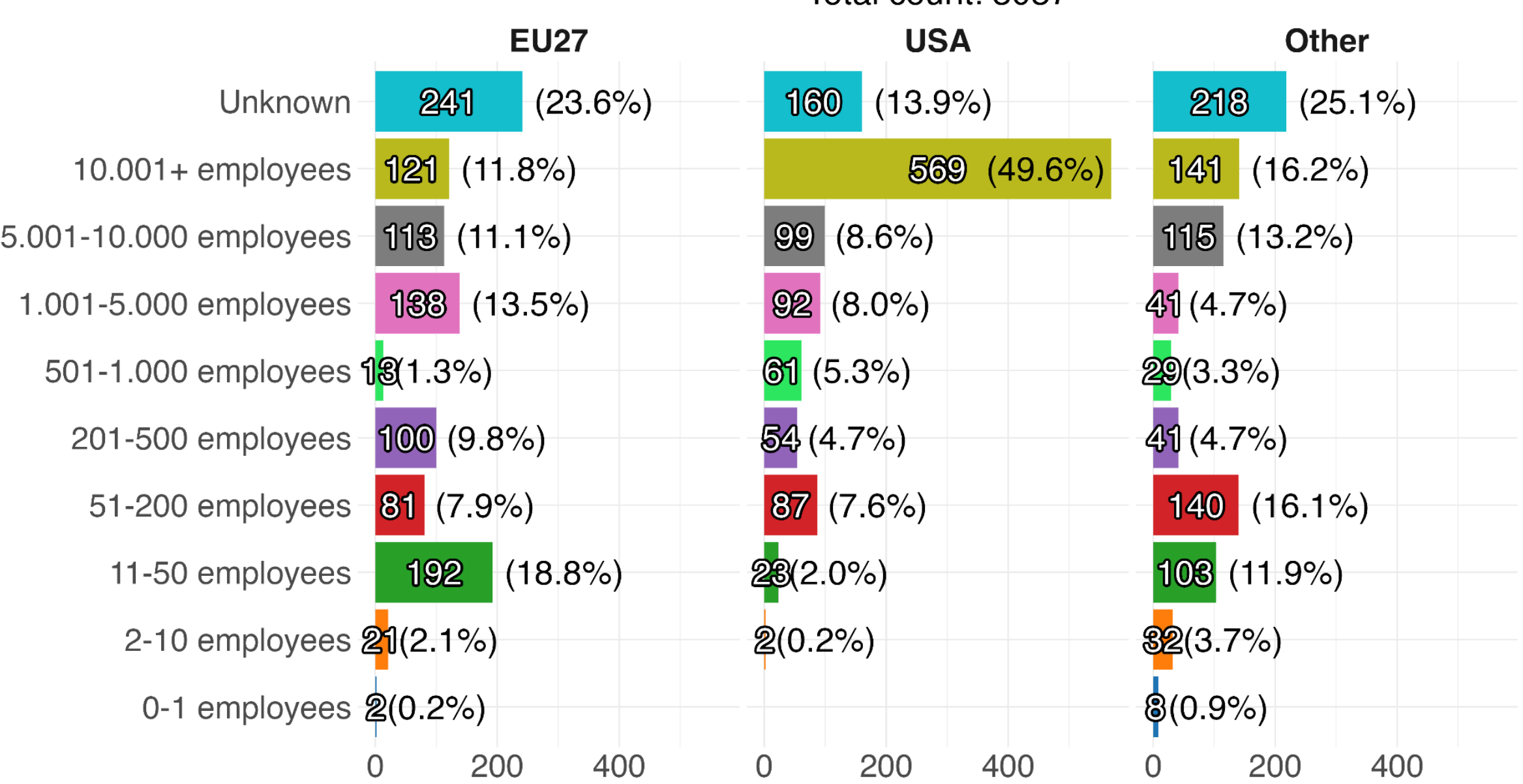

Figure 4. Company size distribution by region. The figure divides the findings of Figure 3 regionally, forming three distinct categories of EU27(left), USA (center) and Other (right).The overall results indicate the presence of larger companies in the US, whereas in the EU, smaller companies are leading towards the hiring of quantum job roles.

We note that the worldwide distribution is largely similar, with one notable exception: The USA hosts a significant preponderance of larger companies (those with 10,000+ employees). These include major technology giants such as IBM, Google, Microsoft, and Nvidia, which are highly active in hiring and building capacity, therefore increasing the count of job posts in the USA for large companies. The United States benefits from significantly higher levels of private and public investment in quantum technologies [44]. Large-scale government initiatives like the National Quantum Initiative Act [40] have attracted substantial private sector funding. Venture capital and corporate R&D funding in the USA have also grown faster than in Europe, fostering an environment conducive both to the growth of QT startups and the intensification of hiring by established companies. On the other hand, the EU27 has a stronger presence of small companies (11-50 employees and 2-10 employees), perhaps owing to the academic strength of the Union which leads in the number of universities and thus resulting in the formation of a number of smaller spin-off companies from research centers. Another area which demonstrates the strength of the corporate sector in the USA is the comparison between academic and corporate job posts in the dataset (Fig. 5). Corporate job roles indicate a greater prevalence in the USA (77.5%) compared to the EU (70.7%).

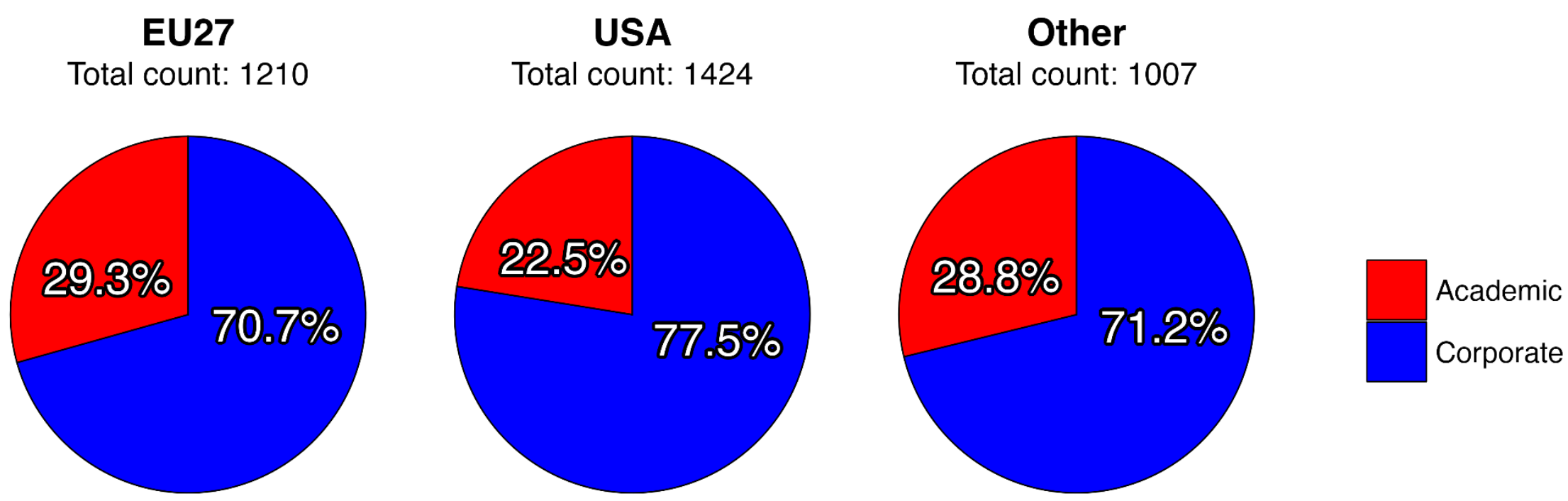

Figure 5. Distribution of job sectors by region.This figure shows how academic and corporate jobs are found, percentage-wise, in each of the three regions: EU27(left), USA (middle) and Other (right). The plots indicate that the dataset is primarily made up of corporate jobs and that furthermore academic jobs are more present in EU27 compared to the USA.

## 4.2 Degree level Requirements

In Fig 6, we show the indicated degree level requirements for the jobs posted worldwide. The overall high demand for PhD candidates (1233 jobs, 34%) may stem from the specialised nature of quantum technologies, which often require advanced research and analytical skills. Companies may prioritize candidates who have conducted extensive research or possess specialized knowledge. It is also notable that when combined together, Bachelor and Master graduates make up 1486 (40.1%) of the total job positions available, indicating they may take less-technical roles, or that companies may offer on-the-job training to bypass the PhD requirement. This may also stem from the increasing number of university programs offering education in QT [7], leading to more available graduates for these roles. Still, the number of jobs requiring a PhD is large in comparison to other industries, as we consider in section 5.

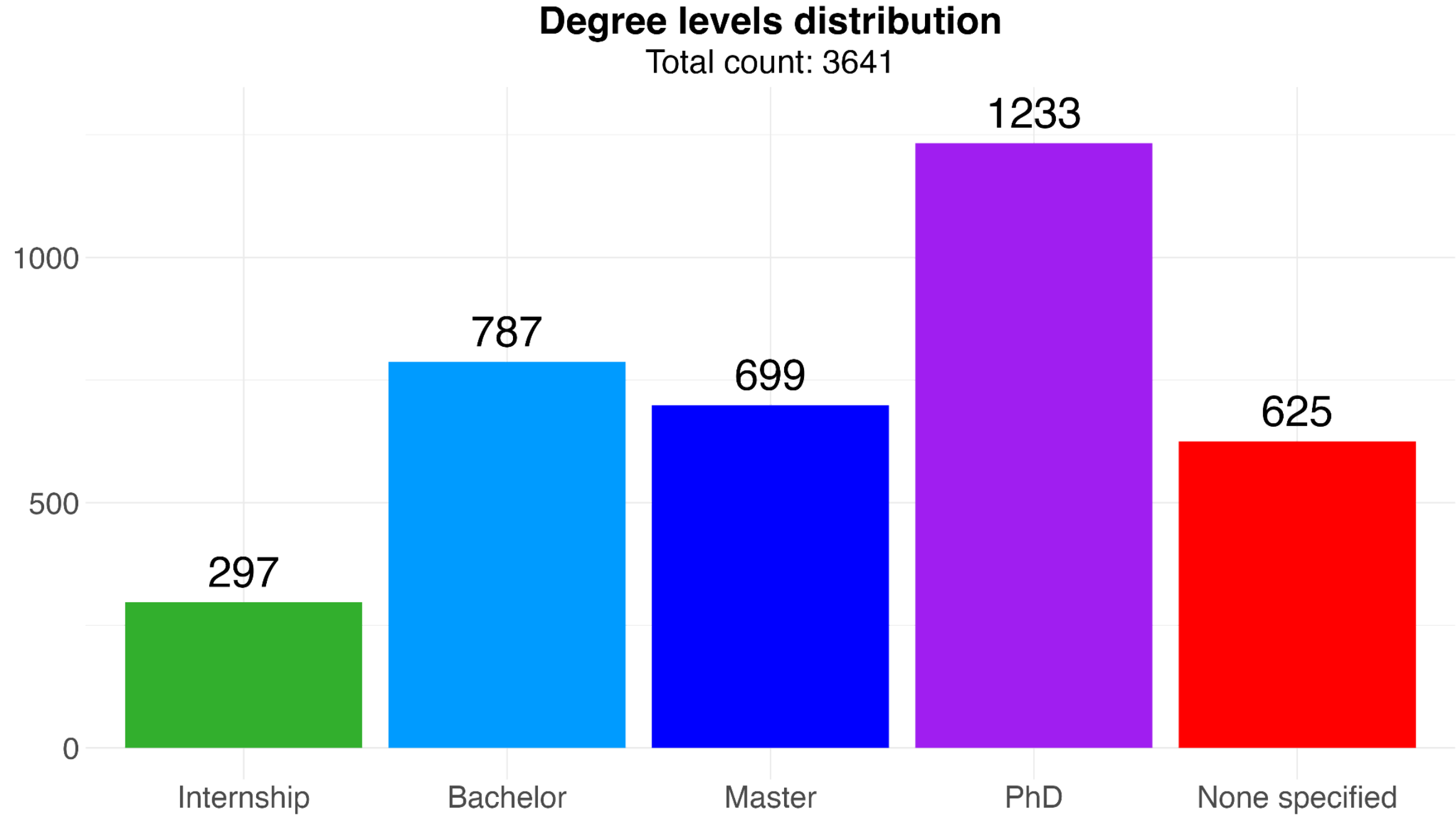

Figure 6. Degree levels requirements of QT jobs. The figure presents the distribution of quantum jobs based on the level of degree they require. The categories are: “Internship”,”Bachelor”, “Master”, “PhD”, and “None specified”.  The resulting distribution shows that while the PhD requirement is still the dominant route to the quantum industry, different degree paths are also available, with a strong presence of Bachelor and Master graduates accepted. Note that 297 job posts represent internships, with no degree requirements.

As far as the regional distribution goes (Fig. 7), we note that the USA has most jobs requiring a PhD (496), and bachelor graduates (426), with a significantly smaller number of roles for master’s graduates (169) as compared to the EU27. This may indicate that the value of a master’s is higher in Europe, potentially relating to the significantly higher number of master’s programs in the EU [7]. Employers could thus be more aware of master’s programs as a viable route to hiring. Furthermore, the educational content of degrees with the same award level is not necessarily the same between countries, and this may influence the hiring practices of local companies.

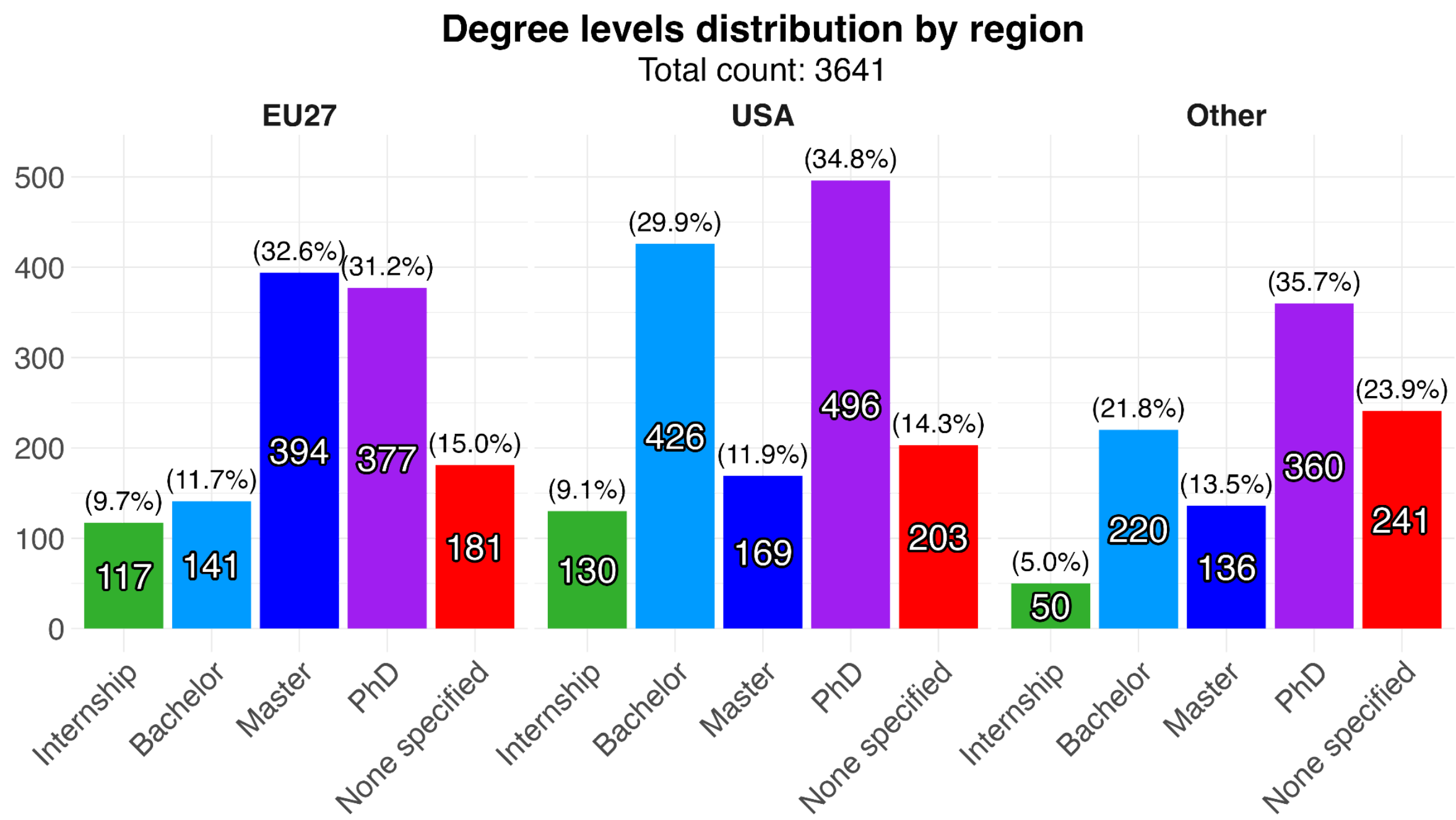

Figure 7. Degree levels requirements distribution by region. This figure divides Figure 6 regionally with the three subfigures representing EU27(left), USA (middle) and Other (right). The results show that the USA hires more Bachelor and PhD graduates, while the EU tends to aim more for Master's graduates.

When comparing among different-sized companies (Fig. 8), what stands out the most, is the overwhelming preference of PhD holders as job candidates for the companies of 1000+ employees. These are likely the tech giants mentioned in subsection 4.1, who require PhD graduates for their research and development roles. Furthermore, as the largest firms, they are likely to offer the most competitive salary and attract the strongest candidates, who are subsequently those with the highest level of qualifications. Smaller companies offer a larger fraction of jobs to candidates with a Bachelor or Master, perhaps suggesting they are open to providing on-the-job training for more technical positions. Companies of all sizes offer a reasonable fraction of their roles, around 10%, as internships potentially aiming at training recent graduates for their needs, in order to take up future job positions in the company.

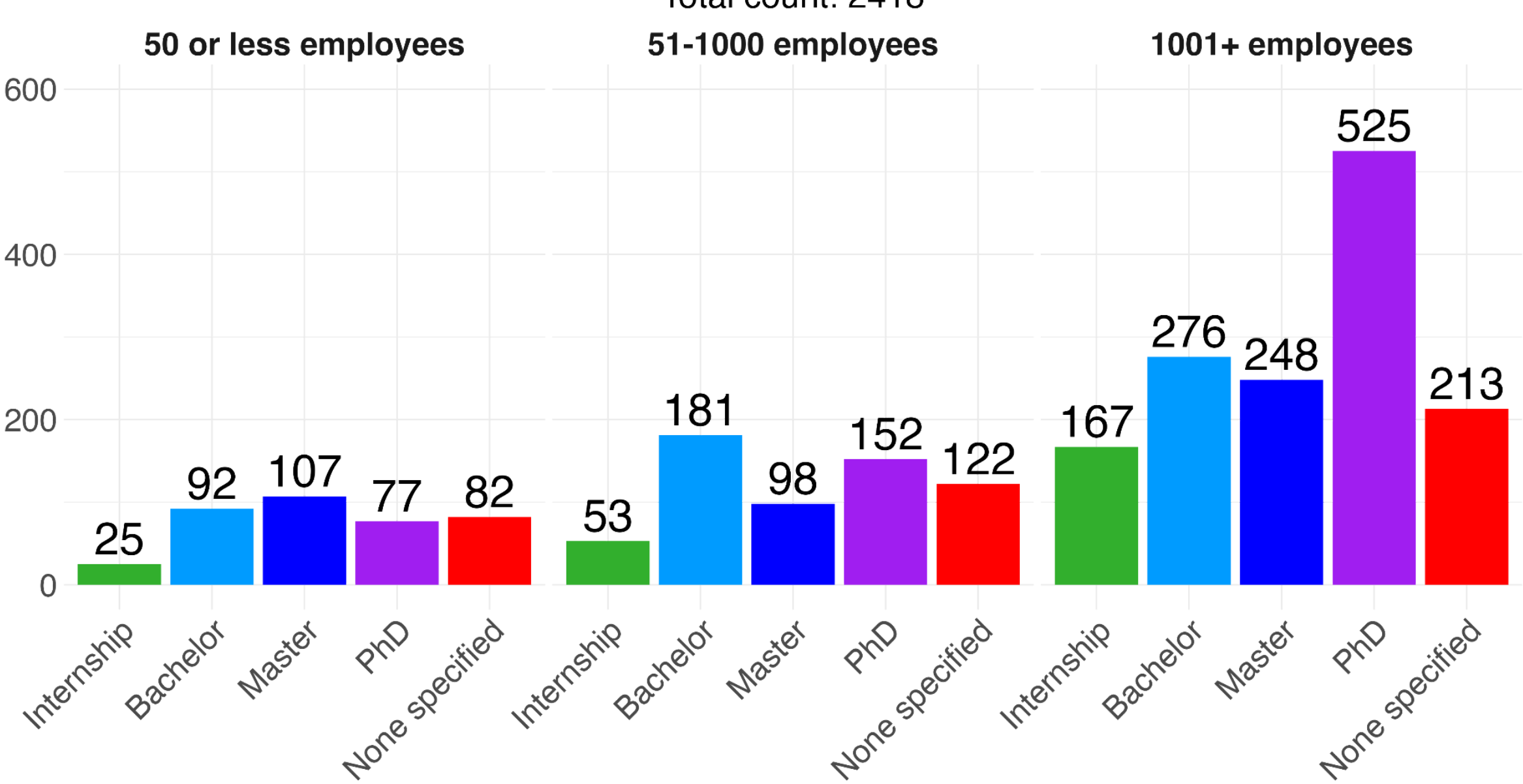

Figure 8. Degree level distribution by company size.This figure divides Figure 6 based on the size of the hiring company: 0-50 employees (left), 51-1000 employees (middle) and 1001+ employees (right). The results show that the USA hires more Bachelor and PhD graduates, while the EU tends to aim more for Master's graduates.

Larger companies have a greater fraction of PhD requiring roles, while there is a more moderate requirement for smaller and medium sized companies.

## 4.3 Degree subject requirements

With regard to the degree subject requirements (Fig. 9), there is a high prevalence of "physics" across all regions (49%), indicating a high need for physicists in the quantum industry. This may stem from the fact that at least up until recent years, physics degrees have provided the most quantum educational content out of all degree programs. Software and electronics/electronic engineering make up the next largest fractions, as it may be the case that more and more quantum course options are available in these degrees [17]. In addition, all of these "hard STEM" degrees offer the same crucial skills such as numeracy and analytical thinking, and these are sought after by all employers in the quantum sector, even aside from the specific quantum knowledge the graduates have.

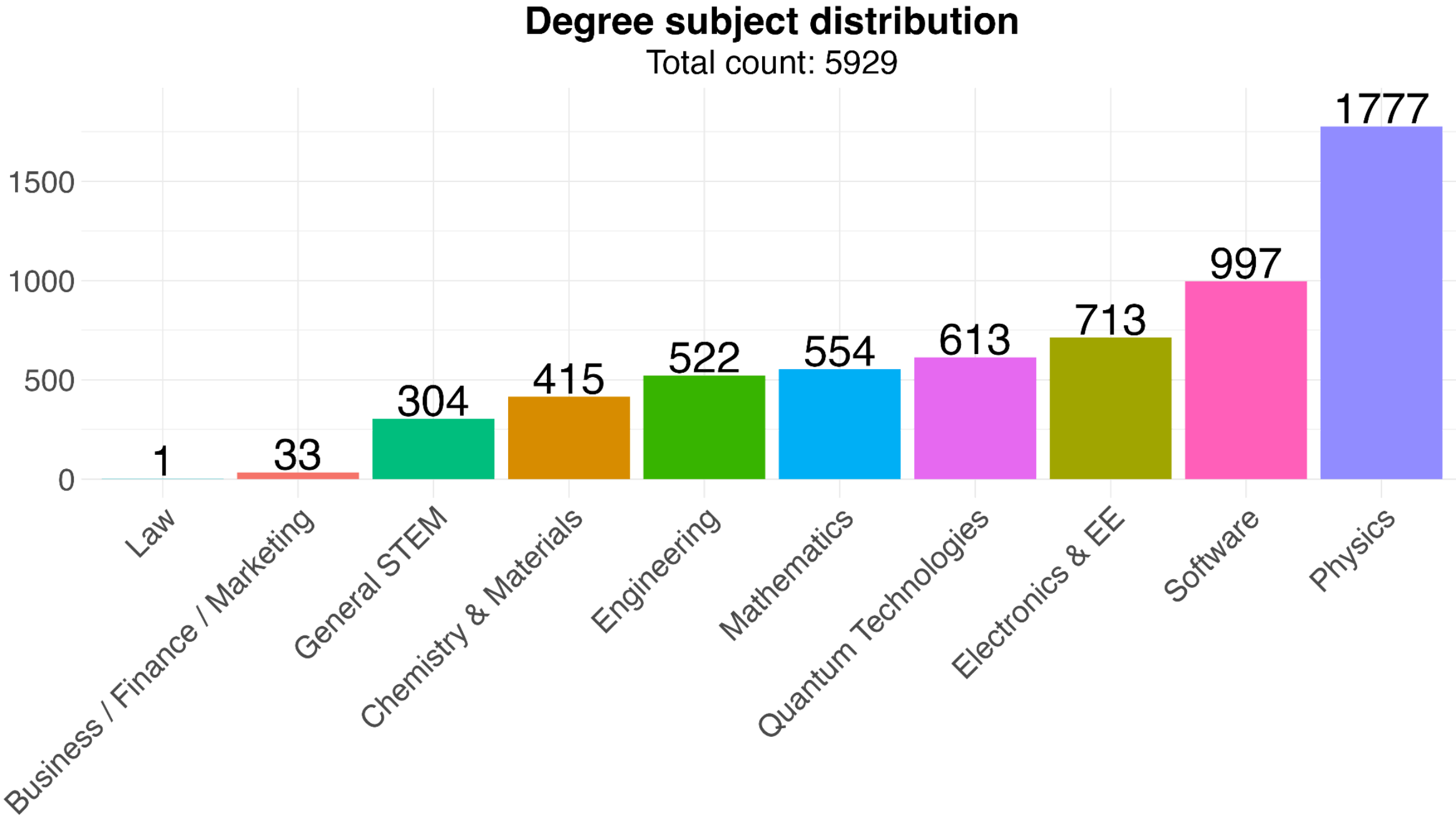

Figure 9. Degree subject requirements of QT jobs worldwide. The figure visualises the frequency of each degree subject found in the different job posts: Physics 49%, Software 27%, Electronics 20% QT 17%, Maths 15%, Engineering 14%, Chemistry & Materials 11%, General STEM 8%, Business /Finance/Marketing 1% / Law 0.03%.

When looking at the comparison between regions regarding the degree subject distribution (Fig. 10), we note that the trends are largely similar between countries, but with a few notable differences. The first is with respect to the "Quantum Technologies" degree demand, which is higher in Europe. As identified previously, this may be due to the high number of specific quantum masters in the region [7]. The United States also show a significant difference to the EU27 regarding hiring electronics/electronic engineering graduates (372 jobs compared to 213), which could suggest a greater readiness in the US quantum industry for applications of quantum technologies, where graduates of electronics and software degrees may have a more prominent role.

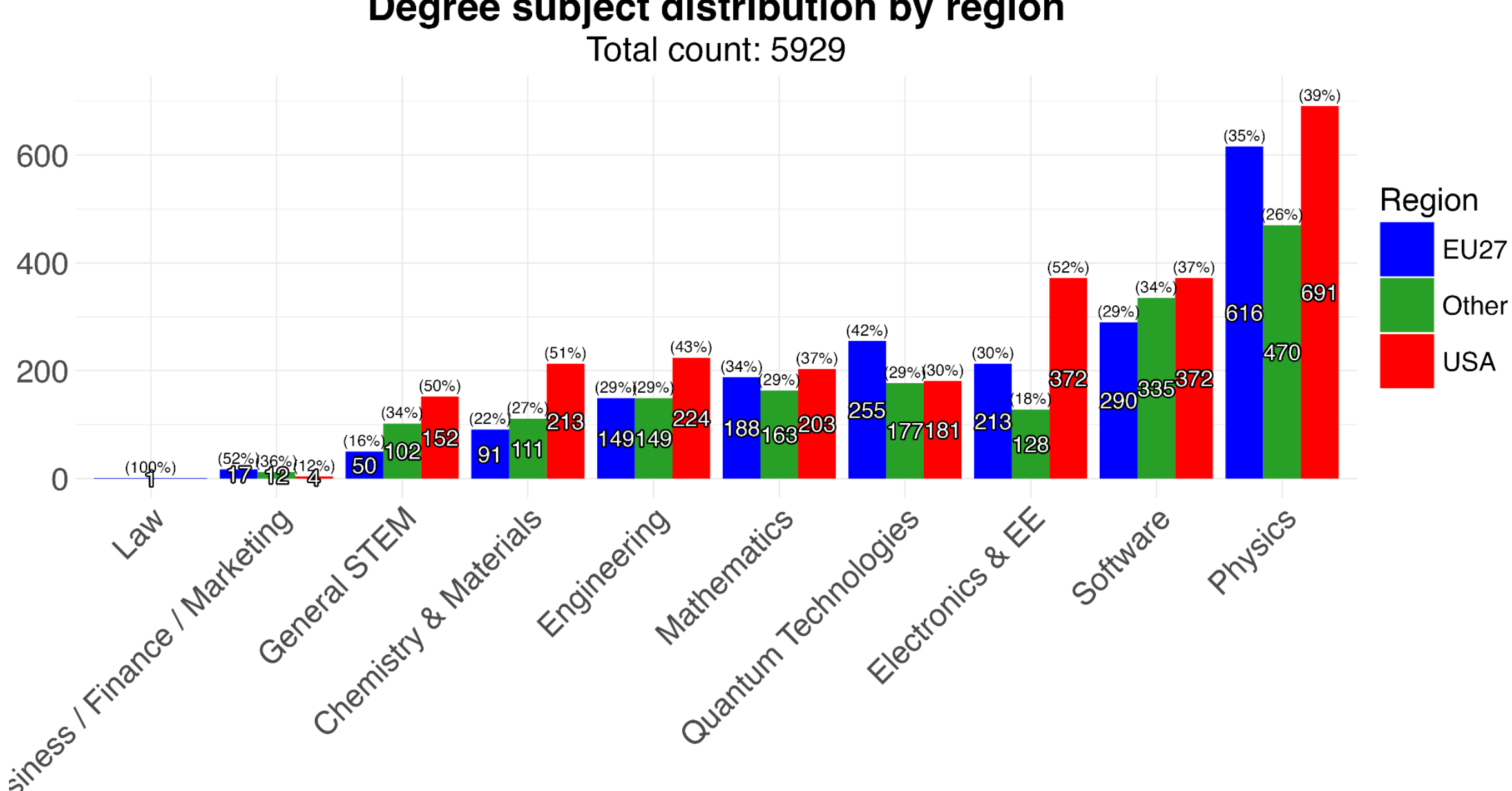

Figure 10. Degree subject requirements of QT jobs worldwide by region.This figure divides Figure 9 regionally with the three colors representing EU27(blue), USA (red) and Other (green). Physics is the most common degree subject required for quantum jobs for all three regional categories.

## 4.5 Job subjects

With regard to the subject area of the QT jobs, in classification step 5 we considered four classes, namely Computing & Simulation, Communication, Sensing, and Hardware. These cover the so-called QT “pillars” as used in the European Quantum Flagship, Quantum computing, simulation, sensing, and communication. In addition, the hardware class is used to distinguish between jobs related to hardware or software for each of the QT pillars. Each job can take one or more of the classes. For example, Quantum computing (software) and quantum computing (hardware) based on the classification in step 5. Quantum Computing (hardware) is the most prevalent combination, representing jobs relating to computation and involving work with hardware. These are followed by Quantum computing (software), and hardware (applications in multiple quantum domains). Quantum communication has more roles (585) than Quantum sensing (317).

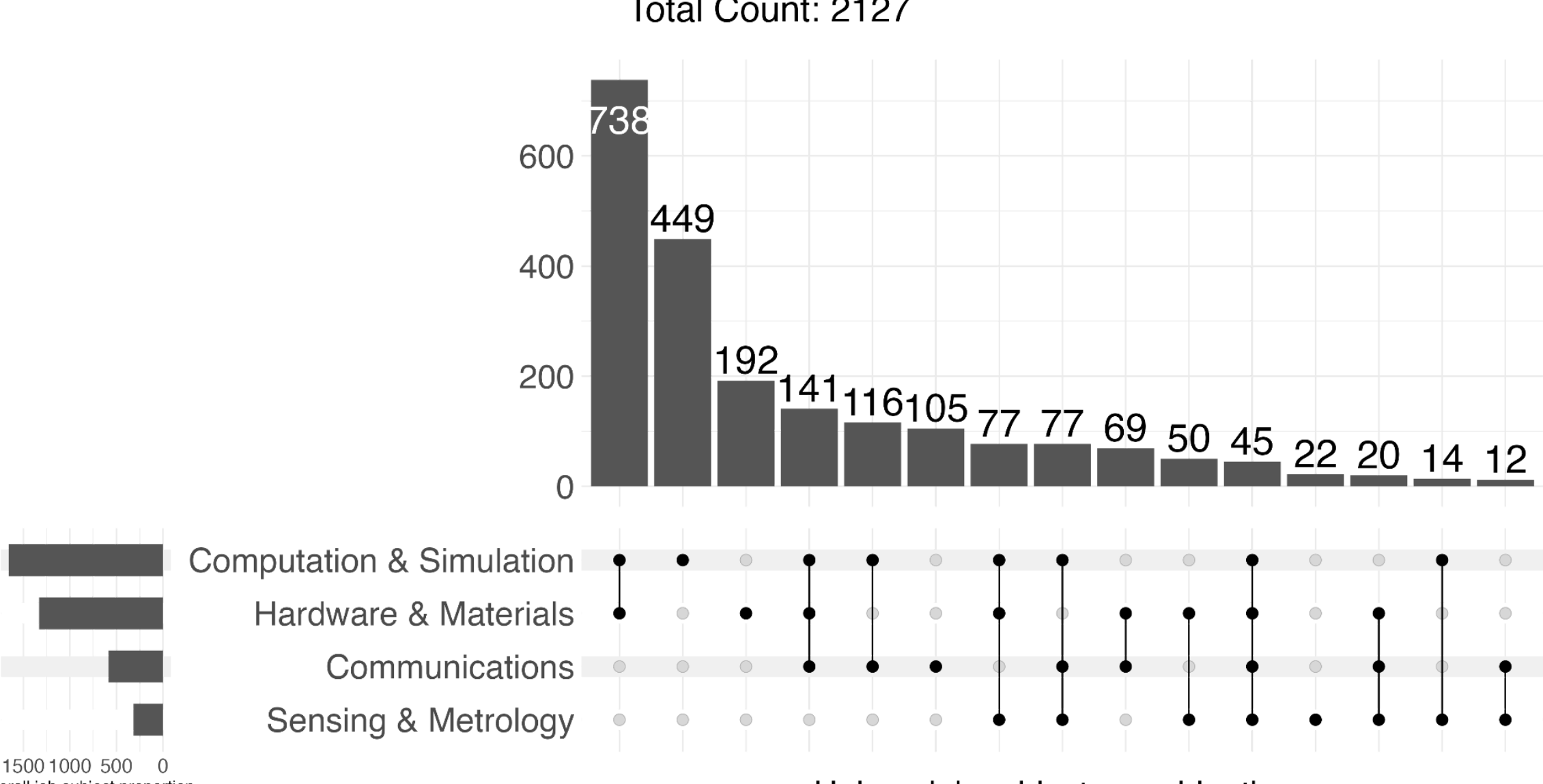

Figure 11. Unique combinations of Quantum domains. Computation & Simulation together with Hardware & Materials can be considered to represent Quantum Computing (Hardware), while Computation & Simulation alone can be representative of Quantum Computing (Software). Quantum Computing seems to be the most represented domain.

It is not surprising that Computing has the most job positions, as the market for quantum computing is estimated at being significantly larger than sensing and communication [45]. The magnitude of the difference is sizeable, however, which can also be observed in Fig.12. Here we see that the EU27 has a greater fraction of job roles in Hardware, and a similar fraction in Computation,while the USA shows a substantial lead in Sensing & Metrology and Communication. This may be related to the size of the companies, as the USA hosts more very large companies, as noted previously. These employers, with the capacity to diversify in different fields, may be the source of the sensing and communication jobs, discussed in section 5.2.

Figure 12: QT domains (Sensing and Metrology, Communications, Hardware and Materials, and Computation and Simulation) compared between EU, USA, and Other regions. Note that this plot counts jobs matching to multiple of the domains more than once - hence the total count of 3891 is above the total number of jobs in the database.

## 4.6 Job roles

The last category of interest is that of the specific roles found in these job postings. The largest class, by a substantial margin, are in technical research and development. R&D scientists or engineers make up 1839 (68.72%) of all corporate roles. By comparison, the other job roles indicated are substantially smaller in magnitude, such that internships (218 jobs) makes up the second-highest number of job posts.This is noteworthy as internships can offer a kind of on-the-job training for graduates who can then go on to work in professional roles. The role *Strategic Planning and Analysis/ Consultant* (138 jobs) encompasses high-level planning, analysis, and strategic development - either internally or as an external service for other companies. This still requires some technical knowledge, and likely a higher degree is preferred. Business development, project managers, and technicians require less QT knowledge, but such positions make up only a small fraction of the available jobs currently. Finally we note that the truly non-technical roles, such as marketing and communication (21 jobs, 0.07%), HR (21 jobs, 0.07%), and financial support (11 jobs, 0.03%), are almost negligible in the QT job market, in the early stages that this paper covers. This truly points towards QT as an infant industry, as we would expect these numbers to be much greater for more established technologies, discussed in section 5.1. The definitions we used for each job role in our analysis, can be found in Appendix 3.

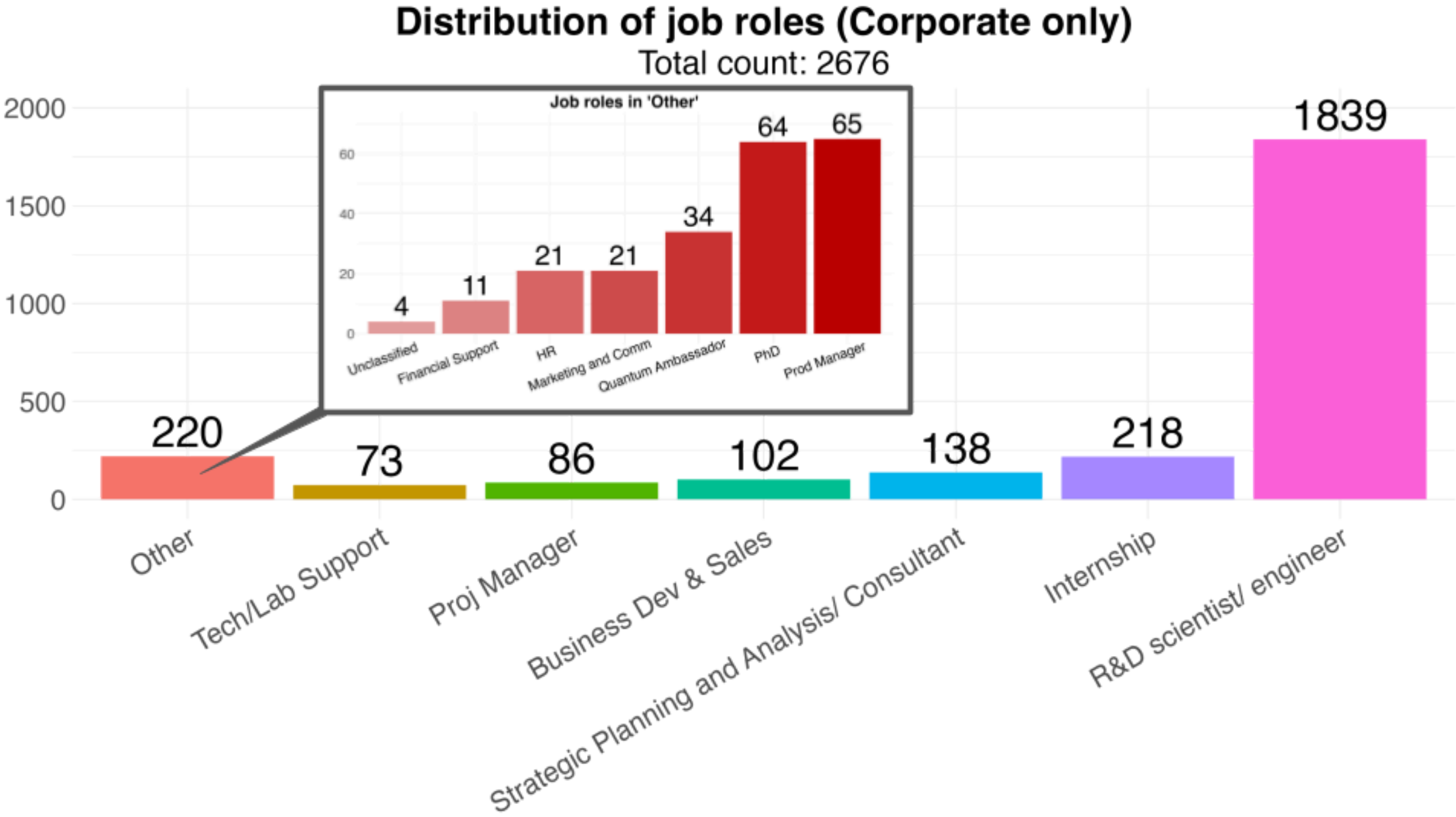

Figure 13. Distribution of job roles that fall under the “Corporate” label. The plot indicates a high dominance of the R&D scientist/engineer while a more moderate distribution, across the rest of the job roles identified, is observed.

## 5. Discussion

Here we consider some of the major implications of these results. Perhaps the most significant is what the trends can tell us about QT as a worldwide field, in terms of its level of technology and market readiness.

### 5.1 QT may still be an immature industry

Several of our results hint at QT being as-yet still nascent in terms of market readiness. When looking at Fig. 3, the distribution of company sizes, we see a significant presence of very large (10,000+ employee) firms and mid-large (5001-10,000) companies, which likely represent the early adopters of QT who can afford to invest in the technology at this early stage of application. Some of the largest companies, by number of jobs posted, include Google, IBM, and Nvidia, which are also some of the largest companies in the world. By comparison, there are relatively few mid-size and small companies in QT, unlike more established fields such as IT services and telecommunications, where the distribution of small, medium, and large firms is more uniform [46].

Fig. 6 demonstrates that there is still a high demand for PhD graduates in the quantum industry, representing over one third of all job posts. This is very high compared to other industries with a greater degree of maturity, such as software development, IT infrastructure, and telecommunications, which have reached a level of commoditization where practical skills outweigh theoretical knowledge [47]. These industries have far more jobs which are either non-technical, or at a lower level of specialisation and hence fewer PhDs are needed. The same trend is observed in the job roles of the quantum industry, where we identified that the significant majority of corporate roles are highly technical. Therefore, companies likely prioritize candidates who have conducted extensive research or possess specialized knowledge. This further highlights the need for specialized master's programs, since companies are possibly looking for experts in the field, while such graduates are a recent phenomenon, as the majority of Master's programs have only been developed within the last 5 years. When taken together, bachelor and master graduates make up 40% of the total job positions available, which is still not insubstantial, and it is likely that they occupy less technical roles, or that companies may offer on-the-job training to bypass the PhD requirement or the lack of specialized knowledge of physics and engineering graduates. As the number of educational opportunities increases, the fraction of graduates able to take up roles in the quantum industry, without the time commitment of a PhD, will likely increase [7].

## 5.2 Quantum computing is the most dominant area, but is it the most ready?

It is noteworthy to compare the subject area, in terms of the QT domains, among the different- sized companies (Fig. 14). It is clear to see that a greater fraction of small companies are working in quantum computing, both in hardware and in software. When comparing quantum computing among small (<50 employees), medium (50-1000 employees) and large (10,001+ employees) companies, we note that it makes up 87%, 80%, and 74% of job roles respectively. This is in line with previous research noting that new startups, in particular, are more likely to favour quantum computing over the other domains[48]. One explanation for this may be that smaller companies are more likely to be spin-outs from academic research labs, with facilities able to develop quantum hardware. Furthermore, smaller companies, due to limited resources, are more likely to focus on a single domains, and quantum computing is perceived as having the greatest potential market value [44]. On the other hand, larger companies may be able to afford to be more diversified in their quantum efforts and therefore, be more likely candidates to include sensing and communication in their product development.

**Distribution of Job Subjects by company size**
Total count across all company sizes: 2408

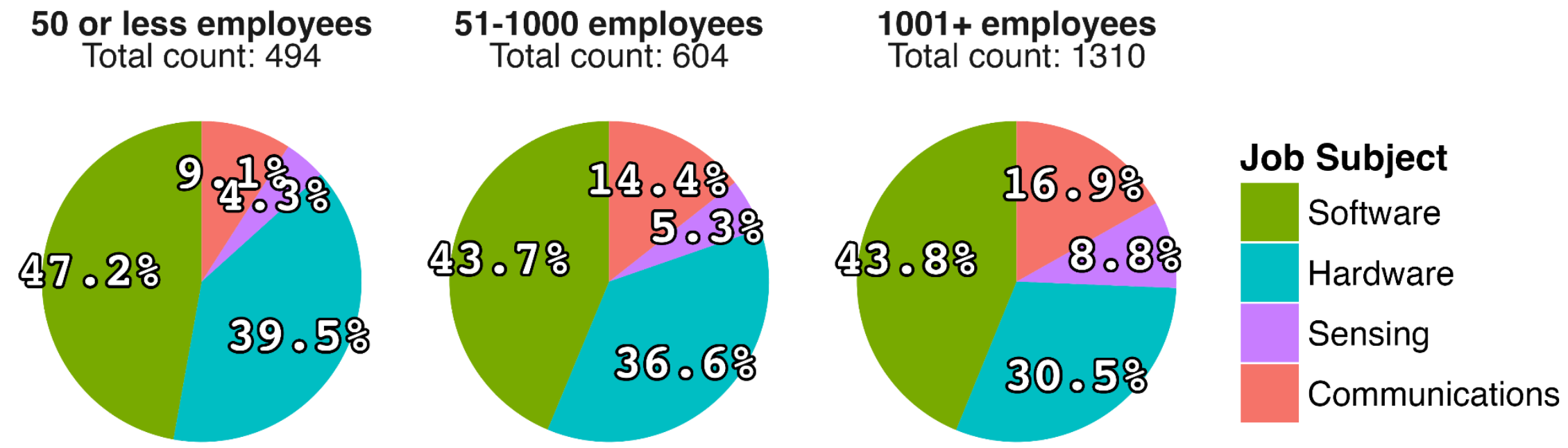

Figure 14. Quantum domains across company sizes. The figure presents the distribution of the four quantum domains across companies with different size: 0-50 employees (left), 51-1000 employees (center) and 1001+ employees (right). Quantum computing (hardware) and quantum computing (software) are among the most expected/needed across all company sizes.

While there is still much discussion about the technology readiness of quantum technologies, and quantum computing in particular [2], it is likely that sensing and communication are in fact closer to the market than computing is. So, while the majority of job roles in the quantum market are addressing computing, with the idea that it may be a more disruptive technology [49], sensing and communication may actually be lower-risk areas to invest in. Our findings around the prevalence of jobs relating to computing also agree with the state of the quantum market, in which computing takes the bulk of total QT revenue among the pillars [59] and computing companies receive the largest share of VC funding in the quantum tech space [60].

## 5.3 The USA leads the market

We have already noted that the largest tech giants are primarily USA-based companies, such as Google, Microsoft, Nvidia, and IBM. When comparing among the top 30 companies posting jobs in the dataset (Fig. 16), 16 (53%) of them are USA companies. Over half of the jobs posted are based in a single country.  Another very telling statistic are the locations of jobs compared to the companies hosting them (Fig. 15). This provides information about how European, American, and other companies are expanding internationally. 79 EU-based jobs are in branches of USA companies (~7.7% of EU jobs), whilst only 15 USA-based jobs are in branches of EU companies (~1.3% of USA jobs), a difference of almost 600%.

**Distribution of job region vs company's headquarters region**

Total count: 3037

*The fractions are the percent of EU27 jobs in USA-based companies and USA jobs in EU27-based companies*

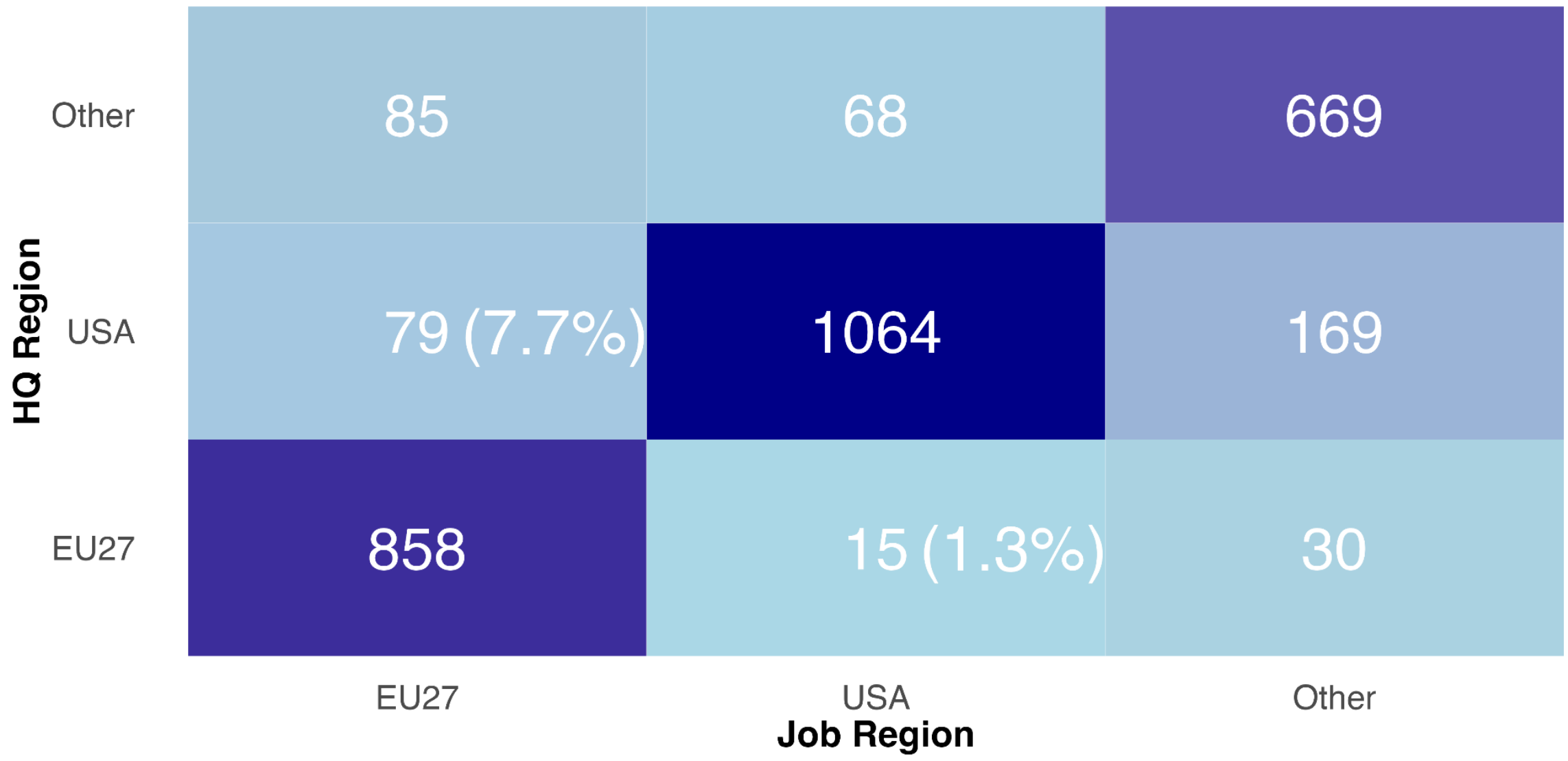

Figure 15. Matrix showing the distribution of job region vs the hiring company's headquarters region. All regions have the most jobs located in their headquarters nation. However, 7.7% of the jobs posted in the EU have headquarters in the USA, while only 1.3% of USA job posts are in EU companies.

It is clear that the quantum industry of the USA is actively establishing a presence in the EU, either to tap into the EU talent pool of graduates, where the EU is academically excelling [7], or to expand global operations. The EU quantum ecosystem has a smaller reciprocal influence in the USA, with fewer EU companies expanding their presence there. One implication of this result is that there is a net flow of talent from the EU to USA. Given the already significant disparity between the market size in the USA and the EU, with the USA far ahead in market value, observed in Fig. 16 and by previous sources [44], it is likely that this gap will continue to increase, as USA companies are able to further their international expansion.

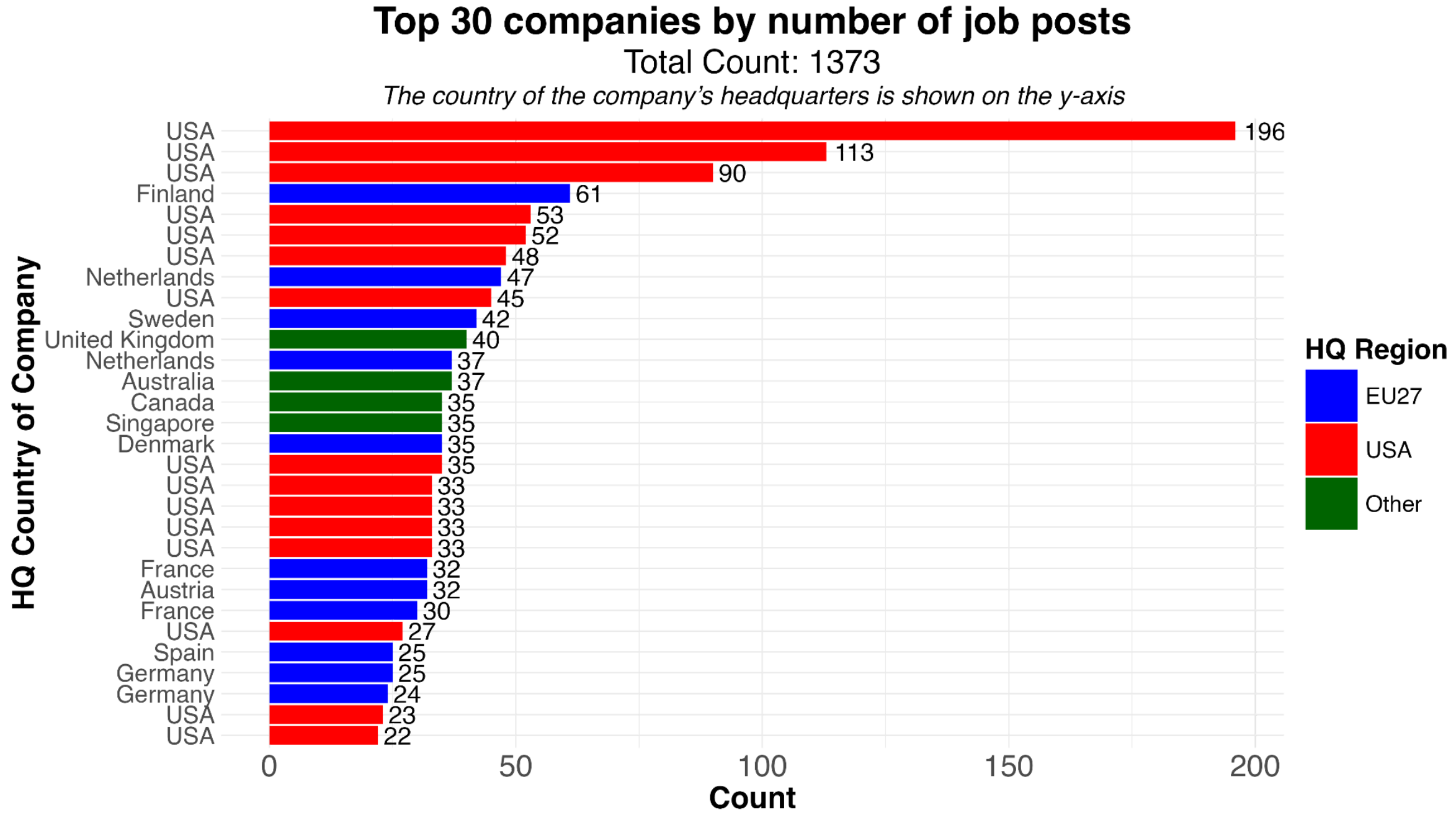

Figure 16. Headquarters (HQ) location of the thirty companies with the highest number of job positions. Note that these companies may have branches in other countries but only the HQ location is shown. It is clear from this figure that there exists a strong presence of large firms in the USA.

## 6. Limitations

Whilst the dataset involved in this research is large, we still acknowledge several limitations on perfect accuracy for job classification. One clear limitation of this research is the data source used, exclusively representing jobs from a single large online job board. However, this website, anonymised to protect personal information, represents one of the largest available worldwide, particularly for technology related jobs. Therefore it seems reasonable to proceed with the dataset for this research, bearing in mind that although it is not comprehensive, it is likely to be reasonably representative of the state of the QT job market, at least in the corporate sector. There is limited information available about how the academic sector uses online job boards, however, and therefore whilst academic jobs are included in our dataset, we consider the academic and corporate roles separately in classification step 4.

Another potential limitation to consider is the relatively small subset of data employed in the ICR assessment (see Appendix 1), which may have influenced the overall reliability of the codebook's quality. While the limited size of the subset could be perceived as a drawback, it actually facilitated quicker iterative loops during the ICR process. This approach enabled more frequent assessments, allowing the team to systematically identify and discuss discrepancies between the annotations of different coders. Through these iterative cycles, valuable insights were gained into areas of inconsistency, which were then used to refine and improve the codebook. Consequently, although the smaller subset may have affected initial reliability metrics, it ultimately contributed to a more robust and thoughtfully developed coding framework.

# 7. Conclusion and outlook

In this work, we have examined the state of the QT job marketplace through job posts worldwide. Information extracted has covered job requirements, topic areas, locations, and company sizes. However, a crucial missing element is the specific skills involved in these jobs. What exactly is their day-to-day content? This research has shed light on the educational levels that quantum jobs require, but what exactly should be the content of bachelor, master, and PhD programs in order to most straightforwardly address the needs of the industry? As of now, this remains unclear. However, the competence framework may provide a means to address it. The job posts collected for this research contain information about skills, tasks, and responsibilities. Therefore, with advances in GPT-based classification, it may be possible to extract and cluster these skills. A first pilot of classifying the job market data using the competence framework (CFQT) is currently underway, and has already provided feedback for the latest version of the framework (3.0). In future work, we intend to comprehensively map the job posts and the regional job markets of the EU, USA, and worldwide, to the competence framework, identify the key requirements for a variety of roles, and perhaps take steps towards a truly industry-driven education landscape.

The quantum job marketplace reveals many signs of quantum technology's nascency as an industry still in early stages of development. The high demand for PhD graduates, low number of startups and mid-sized companies, and a market dominated by large USA companies implies that there is still a way to go and much to be done for QT to reach a stage of maturity that enables it to be more present in our daily lives. Education plays an essential role in providing more trained graduates to fill the at-present rather technical roles, most companies in the QT industry need. This requires not only effective teaching methods at the university level, but also widespread efforts towards inspiration and outreach, so that members of the public and high school students may take a step towards studying for STEM or QT degrees. These graduates form the backbone of the quantum industry, and greater attention must be paid on this need for talent if the quantum industry is to mature to market readiness.

## Declarations

The authors declare no conflict of interest

## Availability of data and materials

Data generated as part of this research is property of the European Quantum Readiness Center. Interest parties may contact the authors to request access.

## Competing Interests

The authors declare no competing interests.

## Funding

The authors acknowledge funding from Horizon Europe for the project Quantum Flagship Coordination Action and Support (QUCATS) under grant agreement ID 101070193.

The authors acknowledge funding from Digital Europe for the project Quantum Technology Courses for Industry (QTINdu) under grant agreement ID 101100757

## Author’s contributions

JS conceptualised and supervised the work. SG managed the scientific team and research planning, and led the writing of the manuscript. BM, ZS, EK contributed to the scientific interpretation of the findings. BM wrote parts of the manuscript. OS led the data science including all development and data analysis. AT supervised the data analysis. All authors reviewed and commented on the manuscript.

## Author’s information

Simon Goorney is a researcher in the European Quantum Readiness Center, based at Aarhus University. He is responsible for implementation of education programs for Quantum Technology, including Europe’s largest QT education project DigiQ. He also studies the changes in the education and industry landscapes which arise from community innovations.

Eleni Karydi is a researcher at the European Quantum Readiness Center, based at Aarhus University. Her research focuses on the Quantum Technology ecosystem in Europe.

Borja Muñoz is a researcher at the Center for Hybrid Intelligence at Aarhus University. He explores the quantum technology ecosystem from a societal perspective, including aspects such as emerging ethical approaches and educational trends.

Otto Santesson is a data scientist and a graduate of the Cognitive Science B.Sc. program at Aarhus University. As part of the European Quantum Readiness Center, they have worked for the past two years on data extraction and analysis of the Quantum Technology job market. They are currently pursuing a degree in Electronic Engineering, with a particular interest in neuromorphic engineering.

Jacob Sherson is Professor of Management in Aarhus University and of Physics at the Niels Bohr Institute, Copenhagen University. He is director of the Center for Hybrid Intelligence and the European Quantum Readiness Center, and a leading figure in educational reform for Quantum Technologies in Europe.

Zeki C. Seskir is a researcher at the Karlsruhe Institute of Technology (KIT) - Institute for Technology Assessment and Systems Analysis (ITAS), in Karlsruhe, Germany, and an EQRC Scholar. He published on topics such as the QT start-up ecosystem, publication and patent landscape, stakeholder engagement, and democratization of QT. His research interests cover a wide range of topics from quantum games to innovation ecosystems.

## Appendix 1: Technical Details

A rigorous validation process is essential for any LLM-based classification to ensure the accuracy and consistency of the output data. The validation process of the pipeline of this paper has been carried out by a team of experts in the QT job market and data scientists, who have iteratively honed each of the prompts of the steps in order to reach a satisfactory classification accuracy. The validation framework utilized in this study is inspired by the works of Shah [50] and Pangakis et al., [51] and has been customized to meet the specific objectives and resource constraints of our project.

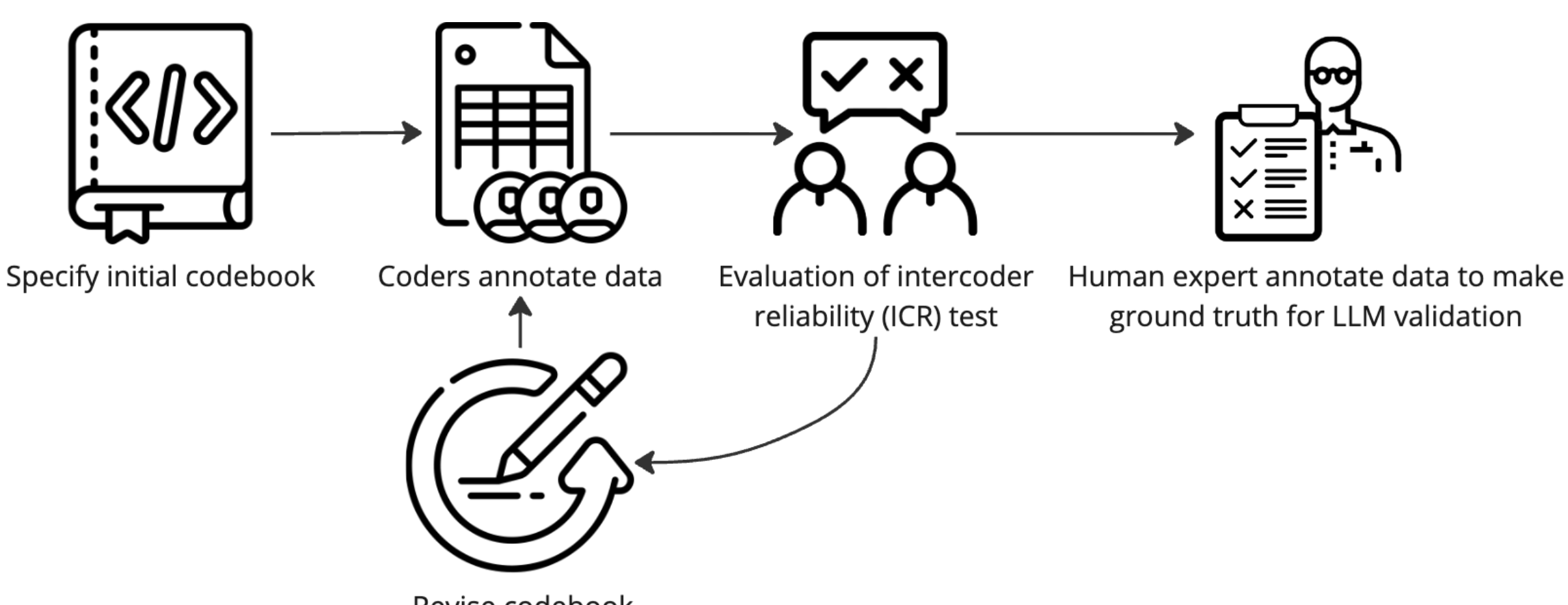

Figure 17. Initial phase of the validation framework. The figure illustrates the different steps involved in composing a codebook which acts as the basis for the classification prompt used. The process is iterative and continues until the codebook has passed an intercoder reliability test.

The validation process of a classification step begins by formulating an initial codebook like any traditional coding task [52-53]. Human coders then annotate the data based on the instructions of the codebook; in our framework, three non-expert coders each had to label a subset of the data (25 job posts) whereafter the labelled data was compared against each other in an intercoder reliability (ICR) test [54]. The purpose of performing the ICR test is to ensure that the codebook is interpreted as intended and to catch any potential edge cases that were not accounted for in the initial codebook specification. Discordant labels formed the basis for revising the codebook, such as using more concise wording or expanding the codebook to encompass previously unforeseen cases. Once the codebook had been revised, the coders would once again label the data so a new ICR test could be conducted. This iterative flow would continue until there was at least 90% agreement between the coders. Once the codebook has passed the test, a human expert (i.e. a person well-acquainted with the QT job market) uses it to annotate 200 job posts in order to create ground truth that can be used in the assessment of the LLM-classification performance. This represents over 5% of the database and is in line with recommendations for LLM-classification validation from prior studies [51]. The next phase of the validation process involves converting the codebook into a prompt for the LLM. Although LLMs are trained on and adept at handling natural language, adjustments are made to ensure the output adheres to a format that the codebase can process correctly.

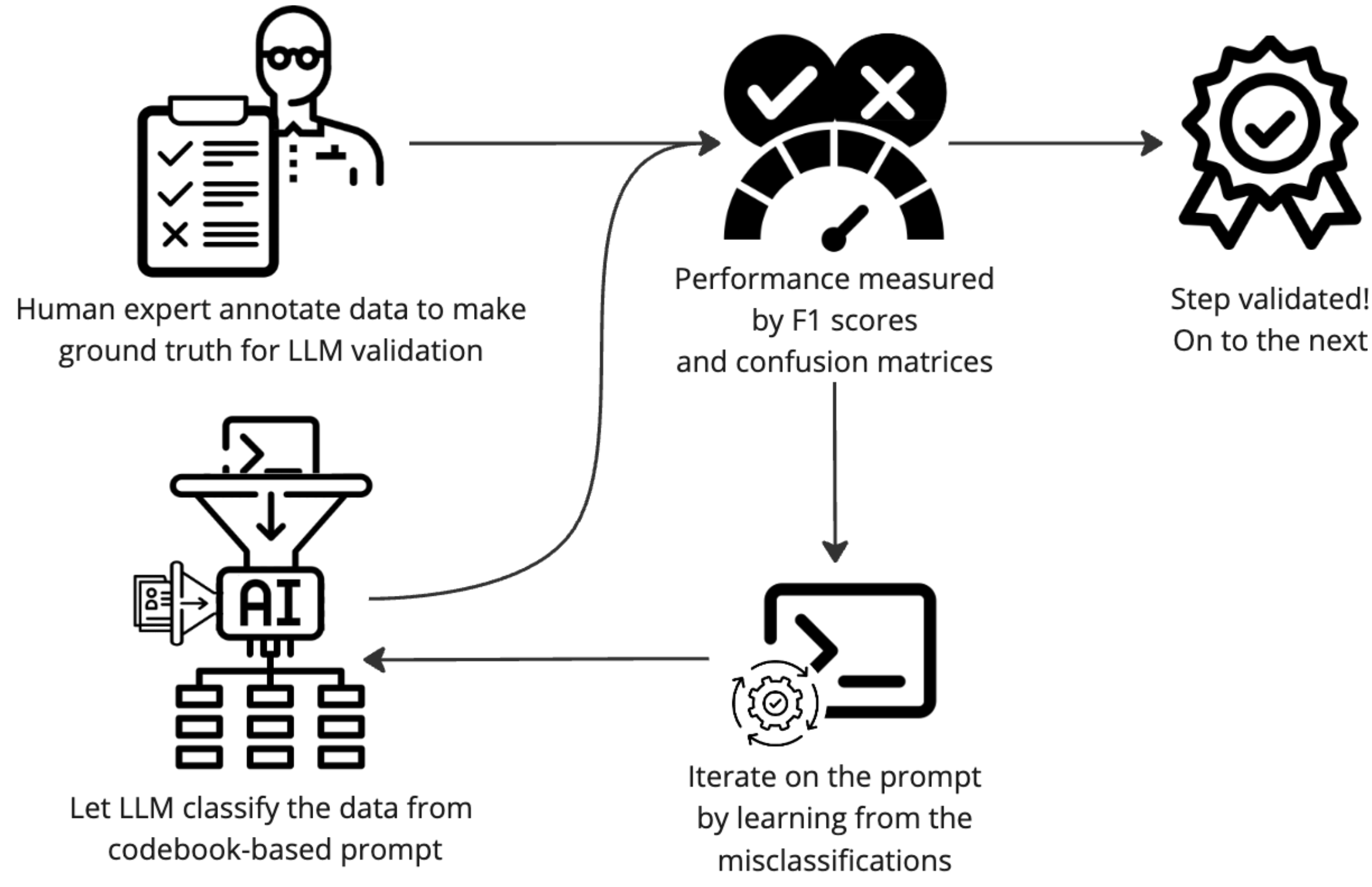

Figure 18. Second phase of the validation framework. The figure shows the process of comparing the ground truth in the form of annotations by a human expert with the classifications of the LLM. If the classification performance is unsatisfactory, the prompt is iterated upon and the LLM classifies the data anew. This is repeated until the performance is adequate.

Once the first version of the prompt for classification has been made, the same 200 job posts, where the true classifications have been defined, are classified by the LLM using the prompt. The classifications made by the LLM are then compared to the ground truth and the performance is evaluated using the weighted average of the F1-scores for the labels along with the associated confusion matrix. The weighted average of the F1-scores was selected as the performance metric for the classification task, as it offers a more robust assessment for datasets with imbalanced label distributions [55]. Since there is no conventional threshold for F1-scores—unlike p-values, which typically indicates significance if below 0.05 [56] — the scores were assessed on an ad hoc basis, using the general rule of thumb that a score of 0.7 or higher is regarded as good [57]. Following this guideline, the prompt was iteratively refined based on insights from misclassifications; the confusion matrix concisely highlighted any potential biases in the classification, thereby aiding in accurately adjusting the prompt. The process involved letting the LLM reclassify the data using the newly refined prompts, and subsequently generating the new F1-score and confusion matrix. This cycle was repeated until the score was deemed satisfactory. Consequently, the lowest weighted average F1-score achieved across all steps was 89%; below is an example of the classification performance for step 4 of the pipeline, on the job posts labeled 'Corporate' in the previous step.

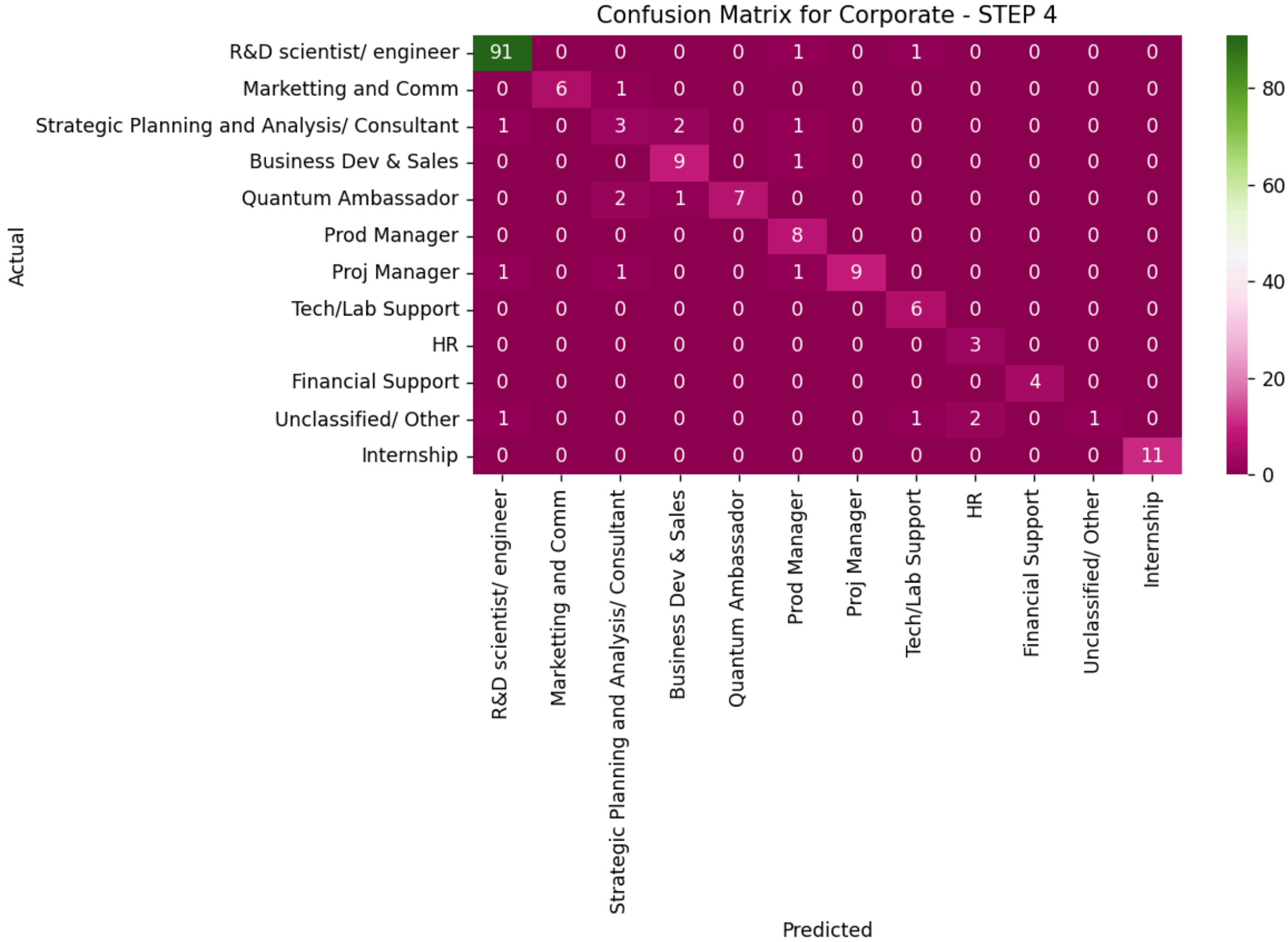

Figure 19. Performance matrix for step 4 of the classification. The confusion matrix provides an overview of the LLM's classification performance across different labels. Notably, the label 'R&D Scientist/Engineer' shows a high proportion of correctly classified job posts, likely due to it being well-defined and easy to delineate. Other roles, such as “Strategic Planning and Analysis / Consultant” may be more difficult to classify accurately, indicated by the off-diagonal components. Overall though, the LLM-classification performs quite well, as seen by the distinctly outlined diagonal of the matrix and quantified by the F1-score.

The majority of the steps of the pipeline were validated using the LLM-model gpt-3.5-turbo-0125 from OpenAI, while the remaining was done using gpt-4o-mini-2024-07-18, also from OpenAI [33]. The models were chosen by balancing cost and performance – the common denominator between the two models is, that they both are the affordable version of their correspondent flagship models. Due to the longevity of the development of the pipeline, gpt-4o-mini was chosen for the later steps, since this model had become available at this point in the process and it was “cheaper, more capable, multimodal, and just as fast as gpt-3.5-turbo” [58].

## Appendix 2: Quantum Job platforms

We have chosen to retrieve all job posts from one platform only, consistently, since that made the process of decomposition easier. In order to decide on the platform we investigated the four most prevalent online platforms for quantum jobs and identified which jobs were repeated across different platforms. As seen in Figure 20, it is clear that one single platform (P2) was demonstrably more comprehensive than the others and so it was chosen as our main job board, representing around 75% of all QT jobs available to view. As a result P2 was used as the data source for this work, which we acknowledge is not a complete source, but it balances proportionally high representation with the development requirements to retrieve jobs.

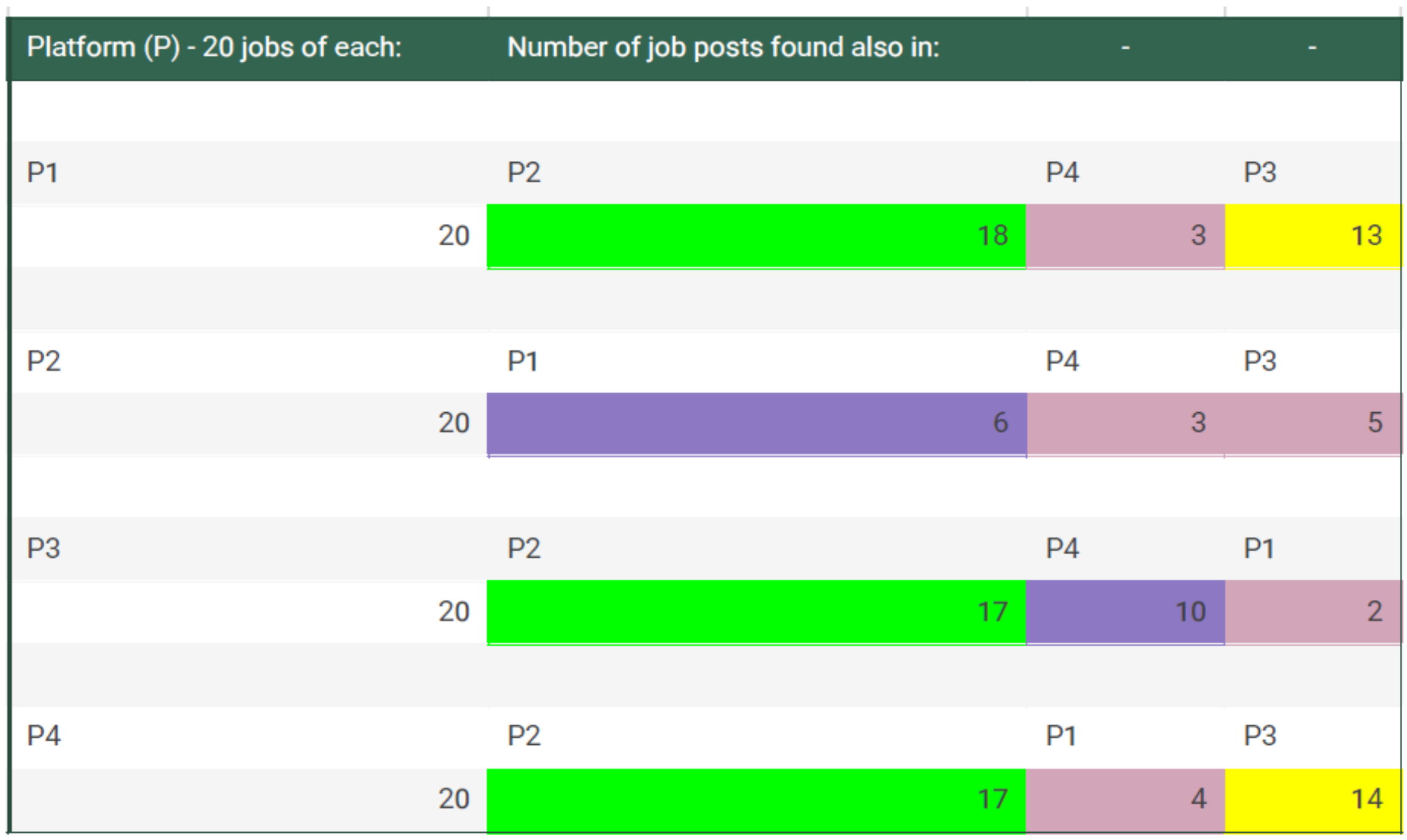

| Platform (P) - 20 jobs of each: | Number of job posts found also in: | - | - |
|---|---|---|---|
| P1 | P2 | P4 | P3 |
| 20 | 18 | 3 | 13 |
| P2 | P1 | P4 | P3 |
| 20 | 6 | 3 | 5 |
| P3 | P2 | P4 | P1 |
| 20 | 17 | 10 | 2 |
| P4 | P2 | P1 | P3 |
| 20 | 17 | 4 | 14 |

Figure 20. Comparison of four different anonymised job platforms. Twenty unique job posts were selected in each platform and then we checked whether or not they had also been posted in each of the other three platforms. The results indicate a clear prevalence of platform P2.

# Appendix 3: Quantum Job Roles

In the fourth step of our classification process we assigned each job post to a specific job role. In order to do that, two human experts in the field went through a large sample of job posts and combining them with their knowledge, created a codebook with broad categories of job roles that can be found in an organization or company engaging in quantum activities. The codebook went through multiple revisions and a few roles that individually represented less than 1% of the total job posts, were togetherc lassified as “Unclassified/ Other”. These broad job role categories are presented below.

First, check if the job is an internship. If yes classify it as “intern”, if no, go to the following categories:

## Academic

### Technical Researcher

- Professor / Tenure-Track Faculty Positions (independent of rank, so includes lecturers, associate and assistant professors and professors)

- Postdoctoral researcher (PhD required)
- PhD student (including industrial PhDs)
- Research assistant (master’s or bachelor’s degree required)
- Engineer

#### Non-Technical Researcher

- Non-technical researcher:  A researcher, of any academic rank (including research assistant, PhD student, Postdocs, and all kind of professorships), whose research topic relates to the Quantum Technologies ecosystem, but is non-technical as they are not involved in development or research of the technology themselves. Example research topics include: Quantum technology education, Quantum technologies impact on society, legal issues around quantum technology development, quantum readiness levels, policies, DEI.

#### Support

- Project Manager: A supporting role responsible for managing academic projects, grants and performing other supporting management tasks in an academic environment.
- Technician & Lab Support: A supporting role who manages equipment and laboratory setup for research staff.
- Communications: A supporting role responsible for creating and implementing the communication strategy of a research group, including event planning, marketing, social media, press releases.
- Educational Developer:A supporting role responsible for the development of educational content around Quantum Technologies, such as curriculum development.
- Financial Support: A supporting role who manages financial issues

#### Unclassified/ Other

### Academic

#### Research and Development

- R&D scientist/engineer: Someone involved in the hands-on technical research, development, designing or building of Quantum Technologies, regardless of rank.

#### Valorization

- Marketing, and Communications: Someone responsible for the marketing department of a company, including Social media, digital marketing, or press releases.
- Strategic Planning and Analysis/ Consultant: This category encompasses roles focused on high-level planning, analysis, and strategy development to guide organizational direction and improve performance. Responsibilities typically include developing business strategies, conducting market and supplier analysis, evaluating policies and regulatory impacts, and supporting product lifecycle management.

- Business Development and Sales: This category includes roles dedicated to expanding business opportunities, driving sales, and fostering relationships with key stakeholders. Responsibilities involve identifying and pursuing new markets and applications, generating leads, negotiating and closing deals, managing sales pipelines, and engaging with customers and partners through various channels.
- Quantum Ambassador: A person in a company whose role is to have a technical knowledge of Quantum Technologies and how it can be applied and adopted by the company, in business development, product development, and stakeholder communications.
- Product manager: A supporting role who is responsible for the strategy, roadmap, and feature definition of a product or product line. PMs work closely with engineering teams, marketing, sales, and customer support to ensure that the product meets market needs and is delivered on time.

#### Support

- Project Manager: A supporting role who manages projects, assisting engineering or research staff in technical and non-technical aspects.
- Technician & Lab Support:  A supporting role who manages equipment and laboratory setup for research or engineering staff.
- Human Resources: A supporting role responsible for finding, screening, recruiting, and training job applicants, as well as administering benefits.
- Financial Support: A supporting role who manages financial issues

#### Unclassified/ Other

## Appendix 4: Prompts

Below we present the prompts that were used for the classification of the job posts, for each separate step.

### Step 1

quantum_related_prompt = """The following is a job post for a position. The job title is: "{}" and the job description is: "{}".

On the basis of the above information, please classify whether the job position is quantum physics or quantum technologies related, or not.
To elaborate, if there is no indication from either the job title, job description, or company description, that the job title is somehow engaged with activities related to quantum physics or quantum technology, then the job position shouldn't be classified as related.
If the job post is quantum physics or quantum technologies related, please return the following string: "True"
If the job post is not quantum physics or quantum technologies related, please return the following string: "False"
Return only the string, without quotation marks, surrounded by square brackets.

decomposition_prompt = """Here is a job offer for a job in the quantum industry: {}

Break down the offer into these 10 sections:
1. Organisation Information: The overview of the recruiting organisation, which often includes their structure, their fields of operations, location, academic ties... If there is no relevant information about the organisation, return "Nothing Explicit" for this field.

2. Job Summary or Overview: A brief description of the main purpose and scope of the job, which often includes the main reason the job exists within the organisation. THIS SECTION DOES NOT INCLUDE GENERAL ORGANISATION INFORMATION, only information specific to this job.

3. Key Responsibilities and Duties: This section lists all the tasks that the job holder will be expected to perform. Be sure to include every task present in the job offer.

4. Required Skills: Specifies the strictly necessary professional skills required to apply for the job. This CAN include technical skills, mastery of certain tools, language proficiency, communication skills, personality traits...

5. Preferred Skills: Specifies the preferred/optional professional skills required to perform the job effectively. This CAN include technical skills, mastery of certain tools, language proficiency, communication skills, personality traits...

6. Required Qualifications: Specifies the strictly necessary educational qualifications required to apply for the job. This CAN ONLY include academic degrees like bachelor's, master's, PhDs, or equivalents from other countries. INCLUDE EVERY ACADEMIC DEGREE FIELD MENTIONED.

7. Preferred Qualifications: Specifies the preferred/optional educational qualifications required to perform the job effectively. This CAN ONLY include academic degrees like bachelor's, master's, PhDs, or equivalents from other countries.

8. Required Previous Experience: Specifies the strictly necessary experience required to apply for the job, as well as the duration associated. This CAN include previous jobs, familiarity with specific tools or technologies...

9. Preferred Previous Experience: Specifies the preferred/optional experience required to perform the job effectively, as well as the duration associated. This CAN include previous jobs, familiarity with specific tools or technologies...

Your output should be a bullet point list. Each entry should have a header beginning with a dash, then the section number, then the section name (without the section description). (e.g: "- 4. Preferred skills and qualifications:"). The header is followed by the corresponding information extracted from the job offer. If the content of a section is empty, write "Nothing Explicit".
All your outputs should be exact segments of the original description. Do not reformulate anything.

ONLY return the list, without anything else.

## Step 2.1

internship_prompt = """The following is a job offer. The job title is: "{}" and the job description is: "{}".
Based on the above information, please classify whether the job position is an internship, studentship etc. or not. A PhD position is not considered an internship, studentship etc., but a regular job position.
If the job post is an internship, studentship etc., please return the following string: "True"
If the job post is not an internship, studentship etc., please return the following string: "False"

Return only the string, without quotation marks, surrounded by square brackets.

## Step 3.1

recruitment_agency_prompt = """Consider the workplace named "{}", with its following description: "{}".
On the basis of the above information, classify whether the organisation is a recruitment company. Return only the label as a Python boolean (True if it is a recruitment company and False if not), surrounded by square brackets.
If there is no explicit information, classify it as False.

## Step 3.2

academic_from_org_info_prompt = """Consider the workplace named "{}", with its following description: "{}".

On the basis of the above information, classify whether the organisation is an academic institution or not.

- A public research organisation (university, public research laboratory) should be classified as "True".
- A private organisation (corporation, firm, private laboratory, spinoffs of universities, recruitment company...) should be classified as "False".

Here are a set of additional guidelines:

- Private research institutions should be classified as "False", since they are not public organisations.
- Mentions of developing or providing "products" or "solutions" should hint towards a for-profit structure, and thus should be classified as "False"
- Simply doing research or joining a research team isn't enough to be classified as "True". The organisation has to be public to be classified as such.
- If there is too little information to be certain, classify it as "False".
- National laboratories are public research institutions and should therefore be classified as "True".

Return the label as a Python boolean ("True" or "False", without quotation marks), surrounded by square brackets, and an explanation for your choice.

## Step 3.3

academic_from_desc_prompt = """Consider an extract of a job offer containing information about the hiring organisation: "{}".

On the basis of the above information, classify whether the organisation is an academic institution or not.

- A public research organisation (university, public research laboratory) should be classified as "True".
- A private organisation (corporation, firm, private laboratory, spinoffs of universities, recruitment company...) should be classified as "False".

Here are a set of additional guidelines:

- Private research institutions should be classified as "False", since they are not public organisations.
- Mentions of developing or providing "products" or "solutions" should hint towards a for-profit structure, and thus should be classified as "False".
- Simply doing research or joining a research team isn't enough to be classified as "True". The organisation has to be public to be classified as such.
- If there is too little information to be certain, classify it as "False".
- National laboratories are public research institutions and should therefore be classified as "True".

Return the label as a Python boolean ("True" or "False", without quotation marks), surrounded by square brackets, and an explanation for your choice.

## Step 4

non_technical_researcher_prompt = """Examine the following job responsibilities/tasks : {}
Here, different jobs, which are related to quantum world but are NON-TECHNICAL as they are not involved in development or research of the technology themselves. :

- Quantum technology education
- Quantum technologies impact on society
- Quantum legal issues around quantum technology development
- Quantum readiness levels, policies, DEI

If the given job responsibilities/tasks are related to one of the above jobs, return Non-technical researcher . Else, return Postdoctoral researcher.
ALWAYS AND ONLY RETURN THE LABEL Non-technical researcher OR Postdoctoral researcher

product_manager_prompt = """ Examine the following job responsibilities/tasks : {}
Below, two managing position related to quantum world :

Product manager

Focus:
As a Product Manager in the quantum realm, you are the visionary driving the development and success of quantum products. Your primary focus is on understanding market needs, defining product strategy, and guiding the roadmap for quantum computing solutions. You collaborate with cross-functional teams—such as R&D, engineering, and marketing—to conceptualize products that meet customer demands and stay ahead of industry trends. Your role is strategic, ensuring that the product not only fits the market but also advances the company's long-term objectives in quantum technology.
Key Responsibilities:

- Define Product Vision and Strategy: Develop a clear vision for quantum products based on market research and customer insights.
- Market Analysis: Identify market opportunities and customer needs within the quantum computing industry.
- Product Roadmapping: Create and manage the product roadmap, prioritizing features and functionalities.
- Cross-Functional Collaboration: Work with engineering and design teams to translate requirements into actionable product specifications.
- Stakeholder Communication: Liaise with sales, marketing, and executive teams to align product goals with business objectives.
- Performance Monitoring: Analyze product performance and customer feedback for continuous improvement.

Project manager
Focus:
As a Project Manager in the quantum sector, your primary responsibility is the successful planning and execution of specific quantum technology projects. You focus on the "how" and "when," ensuring that projects are completed on time, within scope, and within budget. Your role is operational and tactical, involving detailed project planning, resource allocation, and risk management. You coordinate among scientists, engineers, and external partners to keep projects on track and resolve any issues that arise during the project lifecycle.
Key Responsibilities:

- Project Planning: Define project scope, objectives, and deliverables in collaboration with stakeholders.
- Timeline and Budget Management: Develop detailed project schedules and manage budgets effectively.
- Resource Coordination: Allocate tasks and resources efficiently among team members.
- Risk Management: Identify potential project risks and implement mitigation strategies.
- Progress Tracking: Monitor project milestones and adjust plans as necessary to meet objectives.
- Reporting: Provide regular updates to stakeholders on project status, challenges, and successes.

Return the managing position between Project manager and Product manager which fit the best with the given responsibilities
ALWAYS AND ONLY RETURN THE LABEL Project manager OR Product manager

academic_job_role_prompt = """Examine the following job responsibilities/tasks : {}
Below, find different job positions :

Professor
Role: Senior academic with extensive experience, often leading research and academic departments
Responsibilities: Leading research projects, publishing significant papers, mentoring graduate students, and teaching advanced courses.

Associate professor
Role: Mid-level academic, often on the path to becoming a full professor.
Responsibilities: Conducting and leading research, publishing, teaching, and mentoring students, along with some administrative responsibilities.

Assistant professor
Role: Entry-level tenure-track academic position focused on developing research and teaching portfolios.
Responsibilities: Research, publishing, teaching, and supervising students, working towards tenure.

Postdoctoral researcher
Role: Temporary research position post-PhD, focused on further specialization.
Responsibilities: Conducting research, publishing, mentoring PhD students, and possibly teaching.

PhD student
Role: Graduate student working towards a doctoral degree, focusing on original research.
Responsibilities: Research, dissertation writing, coursework, and sometimes teaching or assisting.

Research assistant
Role: Junior researcher assisting with various research tasks, often a PhD or master's student.
Responsibilities: Data collection, analysis, experiment setup, and supporting research publication.

Engineer
Role: Provides technical expertise, often in the design and development of research systems.
Responsibilities: Designing experiments, analyzing technical data, and possibly teaching technical skills.

Non-Technical researcher
Role: Researcher focused on the broader impact of Quantum Technologies without direct involvement in technical development.
Responsibilities: Researching topics like quantum education, societal impact, legal issues, policies, diversity, equity, and inclusion (DEI) in quantum technology.

Project manager
Role: Manages academic projects, including grant management and administrative tasks.
Responsibilities: Coordinating research projects, managing budgets, timelines, and ensuring compliance with grant requirements.

Technician / Lab support
Role: Manages laboratory equipment and setups to support research activities.
Responsibilities: Maintaining lab equipment, setting up experiments, troubleshooting technical issues, and ensuring safety protocols.

IT support
Role: Manages IT infrastructure for research staff.
Responsibilities: Installing, configuring, and troubleshooting computer systems, managing networks, and supporting research software needs.

Communications
Role: Develops and implements communication strategies for research groups.
Responsibilities: Planning events, managing marketing and social media, creating press releases, and promoting research outcomes.

Educational developer
Role: Develops educational content related to Quantum Technologies.
Responsibilities: Creating curricula, designing educational programs, and ensuring content is up-to-date with current research.

Legal support
Role: Manages legal aspects of research, including intellectual property and grant agreements.
Responsibilities: Handling legal issues, ensuring compliance with regulations, managing trademarks, and supporting contract negotiations.

Return the  position along those above which fit the best with the given responsibilities
RETURN ONLY THE LABEL

corporate_job_role_prompt = """Examine the following job responsibilities/tasks : {}

Below, find 14 labels associated to job positions for which we described roles, responsibilities and skills associated to every single label:

R&D Scientist/Engineer
Role and Responsibilities:

- Research and Development: Conduct cutting-edge research in quantum physics, quantum computing, or related fields to advance scientific knowledge.
- Experimentation and Analysis: Design and perform experiments, collect data, and analyze results to validate theories or develop new technologies.
- Innovation: Develop new methodologies, technologies, or products leveraging quantum principles.

- Collaboration: Work with cross-functional teams, including other scientists, engineers, and stakeholders, to achieve research objectives.
- Publication and Presentation: Publish findings in scientific journals and present at conferences to contribute to the scientific community.

Skills:

- Technical Expertise: Deep understanding of quantum mechanics, quantum information science, and related disciplines.
- Analytical Skills: Proficiency in data analysis, statistical methods, and computational modeling.
- Programming Skills: Experience with programming languages and tools used in quantum computing (e.g., Python, Qiskit, Cirq).
- Problem-Solving: Strong ability to tackle complex scientific problems creatively and effectively.
- Communication: Excellent written and verbal communication skills for documenting research and collaborating with team members.

Marketing/Communications

Role and Responsibilities:

- Marketing Strategy: Develop and implement marketing plans to promote quantum technologies or services.
- Content Creation: Create engaging content such as articles, blogs, social media posts, and promotional materials.
- Brand Management: Build and maintain the organization's brand image within the quantum industry.
- Public Relations: Manage media relations, press releases, and coordinate with external agencies.
- Market Research: Analyze market trends and customer needs to inform marketing strategies.

Skills:

- Communication Skills: Exceptional ability to convey complex quantum concepts in an accessible manner.
- Creativity: Innovative approach to developing marketing campaigns and materials.
- Digital Marketing Proficiency: Familiarity with SEO, SEM, social media platforms, and analytics tools.
- Strategic Thinking: Ability to align marketing efforts with business goals.
- Interpersonal Skills: Strong networking and relationship-building abilities.

Strategic planning and analysis/consultant

Role and Responsibilities:

- Strategic Development: Formulate long-term strategies for growth and competitiveness in the quantum sector.
- Market Analysis: Evaluate industry trends, market opportunities, and competitive landscapes.
- Consulting Services: Provide expert advice to clients or internal teams on quantum technology implementation.

- Business Modeling: Develop business models, financial projections, and risk assessments.
- Stakeholder Engagement: Communicate insights and recommendations to executives and stakeholders.

Skills:
- Analytical Skills: Strong ability to interpret data and translate insights into actionable strategies.
- Industry Knowledge: In-depth understanding of the quantum technology landscape and its applications.
- Problem-Solving: Ability to address complex business challenges with innovative solutions.
- Communication: Excellent presentation skills and ability to influence decision-makers.
- Project Management: Proficiency in managing projects and coordinating with cross-functional teams.

Business development and sales
Role and Responsibilities:
- Opportunity Identification: Discover and pursue new business opportunities and partnerships in the quantum industry.
- Relationship Management: Build and maintain relationships with clients, partners, and stakeholders.
- Sales Strategy: Develop and execute sales plans to meet or exceed targets.
- Negotiation: Lead contract negotiations and close deals effectively.
- Market Expansion: Explore new markets and channels for quantum products or services.

Skills:
- Sales Acumen: Proven track record in sales and business development roles.
- Industry Insight: Understanding of quantum technologies and their market potential.
- Communication: Strong interpersonal and persuasive communication skills.
- Negotiation Skills: Ability to negotiate terms and agreements that benefit all parties.
- Self-Motivation: Driven to achieve goals and able to work independently.

Quantum ambassador
Role and Responsibilities:
- Advocacy: Promote the adoption and understanding of quantum technologies to various audiences.
- Outreach: Engage with educational institutions, industry groups, and the public through events and presentations.
- Education: Develop educational materials and programs to explain quantum concepts.
- Representation: Serve as the public face of the organization at conferences, seminars, and media engagements.

Skills:

- Communication Skills: Exceptional ability to articulate complex ideas clearly and compellingly.
- Public Speaking: Confidence and proficiency in presenting to large and diverse audiences.
- Educational Skills: Ability to teach and inspire others about quantum technologies.

Product manager
Role and Responsibilities:

- Product Strategy: Define the vision, roadmap, and requirements for quantum technology products.
- Cross-Functional Leadership: Coordinate with engineering, design, marketing, and sales teams to deliver products.
- Market Analysis: Understand customer needs, market trends, and competitive offerings.
- Product Lifecycle Management: Oversee products from conception through launch and ongoing iterations.
- Stakeholder Communication: Serve as the liaison between technical teams and business stakeholders.

Skills:

- Strategic Thinking: Ability to align product development with business objectives.
- Technical Knowledge: Understanding of quantum technologies and their applications.
- Project Management: Proficiency in managing timelines, resources, and deliverables.
- Communication Skills: Effective at conveying ideas and influencing others.
- Customer Focus: Keen insight into user needs and experience.

Project manager
Role and Responsibilities:

- Project Planning: Define project scope, objectives, timelines, and deliverables related to quantum initiatives.
- Team Coordination: Lead cross-functional teams to ensure project milestones are met.
- Resource Management: Allocate resources efficiently and manage budgets.
- Risk Management: Identify potential risks and implement mitigation strategies.
- Reporting: Provide regular updates to stakeholders on project status and performance.

Skills:

- Organizational Skills: Excellent ability to manage multiple tasks and priorities.
- Leadership: Strong team management and motivational skills.
- Methodological Expertise: Knowledge of project management methodologies (e.g., Agile, Waterfall).
- Problem-Solving: Aptitude for resolving issues that arise during project execution.
- Communication: Clear and effective communication with team members and stakeholders.

Technician / Lab support
Role and Responsibilities:

- Laboratory Maintenance: Set up, calibrate, and maintain laboratory equipment used in quantum research.
- Experiment Support: Assist scientists and engineers in conducting experiments.
- Data Collection: Gather and record experimental data accurately.
- Safety Compliance: Ensure all lab activities adhere to safety protocols and regulations.
- Inventory Management: Keep track of lab supplies and equipment inventory.

Skills:

- Technical Proficiency: Familiarity with laboratory instruments and equipment.
- Attention to Detail: Meticulous in following procedures and recording data.
- Problem-Solving: Ability to troubleshoot equipment issues.
- Teamwork: Collaborative attitude in supporting researchers and colleagues.
- Safety Awareness: Knowledge of lab safety standards and best practices.

IT support

Role and Responsibilities:

- Technical Assistance: Provide support for hardware, software, and network issues.
- System Maintenance: Install and update IT systems used in quantum computing environments.
- Security Management: Implement and monitor cybersecurity measures to protect sensitive data.
- User Training: Educate staff on IT systems and best practices.
- Troubleshooting: Diagnose and resolve technical problems promptly.

Skills:

- Technical Expertise: Strong knowledge of IT infrastructure, including servers, networks, and operating systems.
- Problem-Solving: Ability to quickly identify and fix technical issues.
- Communication: Clear communication with non-technical users.
- Security Awareness: Understanding of cybersecurity principles.
- Customer Service Orientation: Patient and helpful approach to assisting users.

Procurement

Role and Responsibilities:

- Sourcing: Identify and select suppliers for equipment and services needed in quantum projects.
- Negotiation: Negotiate contracts, pricing, and terms with vendors.
- Inventory Control: Manage stock levels to ensure availability of necessary materials.
- Cost Management: Monitor expenditures and seek cost-saving opportunities.
- Compliance: Ensure procurement processes comply with organizational policies and legal requirements.

Skills:

- Negotiation Skills: Strong ability to secure favorable terms.
- Organizational Skills: Efficient in managing multiple procurement activities.
- Industry Knowledge: Understanding of suppliers and products within the quantum technology sector.

- Financial Acumen: Ability to manage budgets and analyze costs.
- Attention to Detail: Thorough in documentation and contract review.

Legal support

Role and Responsibilities:

- Legal Advice: Provide counsel on legal matters related to quantum technologies, such as patents and intellectual property.
- Contract Management: Draft, review, and negotiate contracts and agreements.
- Compliance: Ensure the organization's activities comply with laws and regulations.
- Risk Assessment: Identify legal risks and develop mitigation strategies.
- Dispute Resolution: Handle legal disputes or litigation as needed.

Skills:

- Legal Expertise: Knowledge of technology law, intellectual property, and regulatory compliance.
- Analytical Skills: Ability to interpret complex legal documents and regulations.
- Communication: Clear and persuasive in both written and verbal communication.
- Attention to Detail: Meticulous in reviewing legal documents.
- Problem-Solving: Capable of finding practical legal solutions to business challenges.

Human resources

Role and Responsibilities:

- Talent Acquisition: Recruit and hire qualified candidates for positions within the organization.
- Employee Relations: Manage employee relations, including conflict resolution and performance issues.
- Policy Development: Create and implement HR policies and procedures.
- Training and Development: Coordinate training programs to enhance employee skills.
- Compliance: Ensure adherence to employment laws and regulations.

Skills:

- Interpersonal Skills: Excellent in building relationships and handling sensitive situations.
- Organizational Skills: Efficient in managing HR processes and records.
- Knowledge of HR Laws: Familiarity with labor laws and HR best practices.
- Confidentiality: Ability to handle confidential information discreetly.
- Communication: Strong verbal and written communication skills.

Internal training

Role and Responsibilities:

- Training Needs Assessment: Identify skill gaps and training requirements within the organization.
- Program Development: Design and develop training programs related to quantum technologies and other relevant areas.
- Facilitation: Deliver training sessions and workshops to employees.
- Evaluation: Measure the effectiveness of training programs and make improvements.
- Resource Management: Manage training materials and resources.

Skills:

- Educational Expertise: Knowledge of adult learning principles and instructional design.
- Subject Matter Knowledge: Understanding of quantum technologies to effectively train others.
- Presentation Skills: Engaging and clear in delivering training content.
- Analytical Skills: Ability to assess training outcomes and adjust accordingly.
- Organization: Efficient in planning and scheduling training activities.

Financial support
Role and Responsibilities:

- Financial Planning: Develop budgets and financial forecasts for quantum projects or the organization.
- Accounting: Manage accounting tasks such as bookkeeping, invoicing, and financial reporting.
- Analysis: Analyze financial data to inform strategic decisions.
- Compliance: Ensure financial practices comply with legal requirements and standards.
- Funding Support: Assist in securing funding, grants, or investments.

Skills:

- Financial Acumen: Strong understanding of financial principles and accounting practices.
- Analytical Skills: Ability to interpret financial data and identify trends.
- Attention to Detail: Precise in financial record-keeping and reporting.
- Software Proficiency: Experience with financial software and tools.
- Communication: Clear in explaining financial information to non-financial stakeholders.

Return the position along those above which fit the best with the given responsibilities.
ALWAYS AND ONLY RETURN THE LABEL WITH THE SAME TYPO

## Step 5 - Extraction Substep

pillars_extraction_prompt = """Here is the overview of a job in the quantum industry: "{}".
Here is the list of responsibilities associated to this job: {}.
Identify the scope of the job and list ALL the applications that are mentioned.
Here are a few guidelines for this analysis:

- Examples of items that should be included are general quantum technologies domains, devices that are the object of the job, softwares developed...
- Examples of items that should not be included are skill requirements, working environment, hierarchy... only the technical object of the job matters.
- It does not matter what the focus or the primary aspects of the job are. List everything that is mentioned.
- Try to group items if they are similar enough.
- Contextualize the technologies mentioned by inscribing them in the mentioned application fields.

Return your answer as a single Python list (without any comments, and ensure you only return 1 list), and provide a clear reasoning process.

# Appendix 5: Job Post Samples

In this appendix, we present some samples from the job posts we retrieved. In order to keep them anonymized we have completely removed the "organization description" section and we replaced the name in the description with "XXX".

## Sample 1:

Title: R&D Optoelectronics, HI & Quantum Device Engineer (Experienced) Onsite
Description: XXX is the nation's premier science and engineering lab for national security and technology innovation, with teams of specialists focused on cutting-edge work in a broad array of areas. Some of the main reasons we love our jobs:

- Challenging work with amazing impact that contributes to security, peace, and freedom worldwide
- Extraordinary co-workers
- Some of the best tools, equipment, and research facilities in the world
- Career advancement and enrichment opportunities
- Flexible work arrangements for many positions include 9/80 (work 80 hours every two weeks, with every other Friday off) and 4/10 (work 4 ten-hour days each week) compressed workweeks, part-time work, and telecommuting (a mix of onsite work and working from home)
- Generous vacations, strong medical and other benefits, competitive 401k, learning opportunities, relocation assistance and amenities aimed at creating a solid work/life balance

World-changing Technologies. Life-changing Careers.
These benefits vary by job classification.

What Your Job Will Be Like
We are seeking a creative, motivated individual with a passion for exploring new photonics, optical sensor, and quantum technologies that have high potential to impact national security applications. You will assume a key role within a high performing and cohesive team of applied physicists, electrical and optical engineers, and material scientists to perform R&D that will enable next-generation lasers, sensors, quantum systems and communication links. You will have opportunities to contribute across a full spectrum of relevant technology areas, working from device concept and simulation through fabrication and test. Eventually, you will create and lead original research programs.

On Any Given Day, You May Be Called To

- Perform experimental research involving novel semiconductor-based emitters, detectors, photonic circuits, nanophotonic devices, sensors, quantum systems, MEMs, and infrared imaging arrays.
- Prototype and evaluate photonic microsystems that integrate micro-optics, optoelectronics, and electronics.

- Publish original research in top-tier journals, conference proceedings, and patents.
- Enhance laboratory fabrication, test and prototyping efforts by building new capabilities and supporting existing programs.
- Develop and pursue new opportunities in areas that support the XXX mission by writing proposals and winning funding.

Due to the nature of the work, the selected applicant must be able to work onsite.
The XXX job code title for this position is R&D Electronics Engineer.

Qualifications We Require

- Master's degree in Electrical Engineering, Optical Science, Applied Physics, Physics, Materials Science or a closely related field plus three (3) or more years of relevant technical experience
- Record of publication and innovation, as evidenced by peer-reviewed journal articles, conference proceedings, technical books, or issued patents
- Experience in experimental characterization in optoelectronics, nanophotonics, MEMS, atomic/molecular/optical physics, plasmonics, sensor arrays and/or relevant materials

Experience With One Or More Of The Following

solid-state physics, electromagnetics, quantum or Atomic, Molecular, Optical (AMO) physics, classical optics, optoelectronic device engineering, compound semiconductor materials and/or related fields, ability to acquire and maintain a DOE Q clearance

Qualifications We Desire

- Ph.D in Electrical Engineering, Optical Science, Applied Physics, Physics, Materials Science or a closely related field
- Experience with optical laboratory-based testing using lasers, detectors, modulators, fibers, planar waveguides, and relevant characterization equipment
- Experience With Any Of The Following
  heterogeneous integration and microscale packaging techniques, opto-mechanical design, drawing, and machining, photonic integrated circuits, photon detector focal plane array technology, especially in the infrared, designing and characterizing RF microelectronic circuits and devices, simulation using packages such as COMSOL, ANSYS, Lumerical, or custom software
- Proficiency with software programming and instrument control in languages such as MATLAB, C#, or Python
- Successful project performance on R&D activities, including project scheduling, reporting, and delivering results,
- Experience with proposal writing and desire to establish new research paths through both internal and external customers
- Experience with cleanroom mask/reticle design and related processing techniques such as metal and dielectric materials deposition, wet and dry etching, and lithographic patterning methods

- Ability to work independently and on multi-disciplinary teams in a collaborative environment
- Ability to discover, understand, and articulate customer needs
- Strong oral, written, interpersonal, and organizational skills

About Our Team
The XXX Optoelectronics Department develops compound semiconductor, photonics, and microsystems for national security applications including high-performance computing, communications, renewable energy, sensing and imaging. As part of XXX's XXX Science, Technology and Components Center, we research, design, fabricate, prototype, test and deliver novel lasers, photonic integrated circuits, RF and photonic microsystems, optical MEMS and sensors. We focus on R&D of high performance technology platforms and customized microsystems.

We Use State-of-the-art Materials And Fabrication Capabilities Within XXX To Serve a Broad Mix Of Internal And External Clients. Core Capabilities Include

- Optical, optoelectronic and electronic modeling and simulation
- Epitaxial design, materials growth (MOCVD/MBE) and materials characterization
- Design, microfabrication and testing of III-V photonic devices, RF devices, and infrared detectors
- Heterogeneous microsystem integration and prototyping
- High-speed and harsh-environment optoelectronics, including device reliability assessments

Visit Our Website For More Information

Posting Duration
This posting will be open for application submissions for a minimum of seven (7) calendar days, including the 'posting date'. XXX reserves the right to extend the posting date at any time.

Security Clearance
XXX is required by XXX to conduct a pre-employment drug test and background review that includes checks of personal references, credit, law enforcement records, and employment/education verifications. Applicants for employment need to be able to obtain and maintain a DOE Q-level security clearance, which requires U.S. citizenship. If you hold more than one citizenship (i.e., of the U.S. and another country), your ability to obtain a security clearance may be impacted.

Applicants offered employment with XXX are subject to a federal background investigation to meet the requirements for access to classified information or matter if the duties of the position require a DOE security clearance. Substance abuse or illegal drug use, falsification of information, criminal activity, serious misconduct or other indicators of untrustworthiness can cause a clearance to be denied or terminated by DOE, resulting in the inability to perform the duties assigned and subsequent termination of employment.

EEO

All qualified applicants will receive consideration for employment without regard to race, color, religion, sex, sexual orientation, gender identity, national origin, age, disability, or veteran status and any other protected class under state or federal law.

NNSA Requirements for MedPEDs
If you have a Medical Portable Electronic Device (MedPED), such as a pacemaker, defibrillator, drug-releasing pump, hearing aids, or diagnostic equipment and other equipment for measuring, monitoring, and recording body functions such as heartbeat and brain waves, if employed by XXX you may be required to comply with NNSA security requirements for MedPEDs.

If you have a MedPED and you are selected for an on-site interview at XXX, there may be additional steps necessary to ensure compliance with NNSA security requirements prior to the interview date.

## Sample 2

About the job
XXX is a technology leader in advanced test and measurement instruments based on dynamic signal processing. Our products are used in many challenging research fields by scientists all over the world. XXX's vision is to be the reference for digital instrumentation in leading research and development laboratories.

If you are fascinated by measurement instruments that provide extremely high sensitivity and you are eager to work on modern high-performance applications, take this chance to work in our highly skilled Research and Development (R&D) team.
For our R&D team in XXX we are looking for a Software Engineer.

Responsibilities
- Develop efficient and reliable software for our state-of-the-art instruments;
- Develop software features covering all aspects through concept, implementation, testing and documentation;
- Collaborate effectively in the definition of features, architecture, code reviews;
- Analyze difficult user problems and find effective solutions.

Qualifications
- MSc in Computer Science or relevant engineering field;
- 2+ years of experience in modern C++;
- Good knowledge of data structures and algorithms;
- Fluency in Python;
- Experience in test-driven development;
- Interest in quantum computing control and measurement technologies;
- Strong problem-solving skills, excellent verbal and written communication skills.

We offer a diverse work environment in an international high-tech arena with an open and transparent company culture where personal development forms the basis of our success.

We thrive on cooperation and support distributed decision-making that allows everyone to take responsibility and generate substantial impact from the start and on many levels.
Now is a great time to join the team.
We look forward to receiving your resume and motivation letter.

## Sample 3

About the job
Minimum qualifications:

- Bachelor's degree in Electrical Engineering, Computer Engineering, Physics, a related field, or equivalent practical experience
- 2 years of experience using software for hardware interfacing and data analysis
- 1 year of experience using Python

Preferred qualifications:

- 1 year of experience calibrating qubit systems
- Experience with superconducting qubits
- Experience working in a collaborative software environment (version control, code review, style guides)
- Experience with database engineering and/or data science
- Proficient with Structured Query Language (SQL)

About The Job
The full potential of quantum computing will be unlocked with a large-scale computer capable of complex, error-corrected computations. XXX's mission is to build this computer and unlock solutions to classically intractable problems. Our roadmap is focused on advancing the capabilities of quantum computing and enabling meaningful applications.

As a Quantum Hardware Engineer, you will own and streamline the calibration and characterization procedures for our quantum processors. You will be responsible for ensuring we have an efficient and reliable procedure for device bringup and characterization of large-scale devices such as the XXX processor, as well as fast-turn experimental devices. You will also be responsible for ensuring data from these procedures is easily available for longitudinal data analysis, and work with researchers and engineers to understand the impact of key experiments.

The US base salary range for this full-time position is $119,000-$175,000 + bonus + equity + benefits. Our salary ranges are determined by role, level, and location. The range displayed on each job posting reflects the minimum and maximum target for new hire salaries for the position across all US locations. Within the range, individual pay is determined by work location and additional factors, including job-related skills, experience, and relevant education or training. Your recruiter can share more about the specific salary range for your preferred location during the hiring process.

Please note that the compensation details listed in US role postings reflect the base salary only, and do not include bonus, equity, or benefits. Learn more about benefits at XXX .

Responsibilities

- Develop and maintain an automated calibration system capable of performing initial bringup of a variety of different device architectures with distinct control requirements.
- Improve accuracy and robustness of existing calibration experiments.
- Work with researchers to prototype novel control/calibration techniques towards productionization.
- Calibrate and characterize superconducting qubit devices (large and small scale).
- Develop software tools for longitudinal data analysis and visualization.

XXX is proud to be an equal opportunity workplace and is an affirmative action employer. We are committed to equal employment opportunity regardless of race, color, ancestry, religion, sex, national origin, sexual orientation, age, citizenship, marital status, disability, gender identity or Veteran status. We also consider qualified applicants regardless of criminal histories, consistent with legal requirements. See also XXX's EEO Policy and EEO is the Law. If you have a disability or special need that requires accommodation, please let us know by completing our Accommodations for Applicants form .